\documentclass{aastex61}

\newcommand{\nitii}{{\rm N{\textsc{ii}}\,}}

\newcommand{\neiii}{{\rm Ne{\textsc{iii}}\,}}

\newcommand{\nevi}{{\rm Ne{\textsc{vi}}\,}}

\newcommand{\mgv}{{\rm Mg{\textsc{v}}\,}}

\newcommand{\mgvii}{{\rm Mg{\textsc{vii}}\,}}

\newcommand{\sulii}{{\rm S{\textsc{ii}}\,}}
\newcommand{\suliii}{{\rm S{\textsc{iii}}\,}}

\newcommand{\fevii}{{\rm Fe{\textsc{vii}}\,}}

\newcommand{\fex}{{\rm Fe{\textsc{x}}\,}}

\newcommand{\fexiii}{{\rm Fe{\textsc{xiii}}\,}}

\newcommand{\fexix}{{\rm Fe{\textsc{xix}}\,}}

\received{Month Day, Year}
\revised{Month Day, 2017}
\accepted{Month Day, 2017}
\submitjournal{ApJS}

\shorttitle{Non-relativistic free-free emission due to $n-$distributed electrons}
\shortauthors{de Avillez \& Breitschwerdt}

\begin{document}

\title{Non-relativistic free-free emission due to $n-$distribution of electrons \\ Radiative cooling and thermally averaged 
	and total Gaunt factors}

\correspondingauthor{Miguel A. de Avillez}
\email{mavillez@galaxy.lca.uevora.pt}

\author{Miguel A. de Avillez}
\affil{Department of Mathematics, University of \'Evora, R. Rom\~ao Ramalho 59, 7000 \'Evora, Portugal}
\affiliation{Zentrum f\"ur Astronomie und Astrophysik, Technische Universit\"at Berlin, Hardenbergstrasse 36, 
	D-10623 Berlin, Germany}

\author{Dieter Breitschwerdt}
\affiliation{Zentrum f\"ur Astronomie und Astrophysik, Technische Universit\"at Berlin, Hardenbergstrasse 36, 
	D-10623 Berlin, Germany}




\begin{abstract}
	
	Tracking the thermal evolution of plasmas, characterized by an $n-$distribution, using numerical simulations, requires the 
	determination of the  emission spectra and of the radiative losses due to free-free emission from the corresponding 
	temperature averaged and total Gaunt factors. Detailed calculations of the latter are presented, associated to 
	$n-$distributed electrons with the parameter $n$ ranging from 1 (corresponding to the Maxwell-Boltzmann distribution) to 
	100. The temperature averaged and total Gaunt factors, with decreasing $n$ tend to those obtained with the 
	Maxwell-Boltzmann distribution. 
	
	Radiative losses due to free-free emission in a plasma evolving under collisional ionization equilibrium conditions and 
	composed 
	by H, He, C, N, O, Ne, Mg, Si, S, and Fe ions, are presented. These losses decrease with the decrease in the parameter $n$ 
	reaching a minimum when $n=1$, and, thus converging to the losses of a thermal plasma.
	
	Tables of the thermal averaged and total Gaunt factors calculated for $n$ distributions and a wide range electron and 
	photon energies are presented. 
	
\end{abstract}


\section{Introduction}

Non-thermal electron distributions, e.g., $\kappa$ \citep{vasyliunas1968}, $n$ \citep{hares1979, seely1987}, 
depleted high energy tail \citep{druyvesteyn1930, behringer1994}, and hybrid Maxwell-Boltzmann/power-law tail 
\citep{berezhko1999, porquet2001, dzifcakova2011}, in the low density plasma can occur frequently. 
In principle, this may arise in any place where a high temperature or density gradient exists, or when energy is deposited 
into the tail of the distribution at a rate that is sufficiently high to overcome the establishment of thermal equilibrium 
described by the Maxwell-Boltzmann distribution (hereafter denoted by MB). 

In the astrophysical context $\kappa$- and $n$-distributions have been used to describe the evolution of electrons, e.g. 
during a solar flare event, and to interpret spectral lines. It is possible that microscopic instabilities like a two-stream or 
Farley-Buneman instability \citep{buneman1963, farley1963} can occur under these conditions, which drive the electron 
distribution strongly into non-equilibrium \citep{karlicky2012}. Events of magnetic reconnection, in which topological field 
changes force magnetic energy to be released impulsively and dissipated into heat, accompanied by particle acceleration and 
steep gradients in temperature, should be ubiquitous in the interstellar medium (ISM). However, they will be difficult to 
observe individually due to the small extension of the region in which they happen. Hence observed free-free emission 
might often contain a contribution from $n-$distribution electrons.  

The emission spectra as well as the radiative losses due to free-free emission by an astrophysical plasma  are calculated by 
means of the temperature-averaged and total free-free Gaunt factor\footnote{The Gaunt factor is a measure of the quantum 
mechanical 
correction to the semiclassical cross section of \citet{kramers1923}.} (see discussions in, e.g., \citet[hereafter denoted by 
KL1961]{KL1961}, \citet{carson1988}, \citet{hummer1988}, \citet{janicki1990}, \citet{sutherland1998}, \citet{nozawa1998}, 
\citet{itoh2000}, \citet{vanhoof2014}, \citet[hereafter denoted by AB2015]{avillez2015}  and references therein). 

Here we continue our previous work (AB2015) by carrying out detailed calculations (and produce lookup tables to be used 
in any emission software through interpolation) of the non-relativistic temperature averaged and total free-free Gaunt 
factors  for the interaction of electrons (having an $n$-distribution) with a Coulomb field, which is found, for instance, 
when electrons interact with ions in an astrophysical plasma. 

The structure of this paper is as follows: Section 2 presents the theoretical derivation of the temperature averaged and total 
free-free Gaunt factors. Section 3 deals with the methods and results of the present calculations for $n$-distributed 
electrons. In Section 4 a detailed comparison between our results and previously published by other authors is made. 
Section 5 describes the tabulated data, and the paper is concluded (Section 6) with some final remarks.

\section{Temperature averaged and total Gaunt factor}

The energy emitted per unit time and unit volume by free-free emission from electrons with an energy distribution 
$f(E)$ is given by (see, e.g., KL1961)
\begin{equation}
\label{eq1}
\frac{d\Lambda_{_{\rm ff}}}{d \nu}=\frac{8 \pi^{2}}{c^{3}}\left(\frac{2}{3 
	m_{_{e}}}\right)^{3/2} e^{6} z^{2}n_{e} 
n_{Z,z}\displaystyle \int_{E_{o}}^{+\infty}\frac{1}{E^{1/2}}f(E) \, g_{_{\rm ff}}\left(\epsilon_{i}=\frac{E}{z^{2} 
	\mbox{Ry}},\epsilon_{f}=\frac{E-h\nu}{z^2\mbox{Ry}}\right) dE
\end{equation}
where $\nu$ is the frequency of the emitted photon, $E_{o}=h\nu $ is the minimum allowed energy of the electrons in 
order to emit a photon of frequency $\nu$, $m_e$ the electron mass, $n_{e}$ is the electron density, $n_{_{Z,z}}$ is 
the number density of ion with atomic number $Z$ and ionization stage $z$, and Ry denotes the Rydberg constant.

The integral in the RHS of (\ref{eq1}) is the thermally averaged Gaunt factor, denoted by $\langle g_{_{\rm 
ff}}(\gamma^{2},u)\rangle_{em}$ (with $u=h\nu/k_{B}T$ and $\gamma^{2}=z^{2}\mbox{Ry}/k_{B} T$), multiplied by a normalization 
constant $N_{em}$ defined such that $\langle g_{_{\rm ff}}(\gamma^{2},u)\rangle_{em}=1$ when $g_{_{\rm ff}}=1$. Here $T$ is 
the temperature and $k_{B}$ is the Boltzmann constant. 

With a suitable change of variables ($x=E/k_{B}T$ and $y=x-u$) the thermally averaged Gaunt factor can be written as 
\begin{equation}
\label{new1}
\langle g_{_{\rm ff}}(\gamma^{2},u)\rangle_{em}=
\frac{(k_{B}T)^{1/2}}{N_{em}}\int_{0}^{+\infty}\frac{1}{(y+u)^{1/2}}f\left[(y+u)k_{B}T\right]  
g_{_{\rm ff}}\left(\epsilon_{i}=\frac{y+u}{\gamma^{2}},\epsilon_{f}=\frac{y}{\gamma^2}\right) dy.
\end{equation}

\subsection{Maxwellian-Boltzmann  distribution of electrons}
For the Maxwell-Boltzmann distribution of electrons (\ref{new1}) becomes
\begin{equation}
\langle g_{_{\rm ff}}(\gamma^{2},u)\rangle_{em}=\frac{1}{N_{em}}\frac{2 e^{-u}}{\sqrt{\pi} (k_{B}T)^{1/2}} \int_{0}^{+\infty} e^{-y} 
g_{_{\rm ff}}\left(\epsilon_{i}=\frac{y+u}{\gamma^{2}},\epsilon_{f}=\frac{y}{\gamma^{2}}\right) dy.
\end{equation}
The boundary condition referred above imposes that $\displaystyle N_{em}=\frac{2 e^{-u}}{\sqrt{\pi} (k_{B}T)^{1/2}}$ and 
therefore 
\begin{equation}
\label{eq5}
\langle g_{_{\rm ff}}(\gamma^{2},u)\rangle_{em}=\int_{0}^{+\infty} e^{-y} 
g_{_{\rm ff}}\left(\epsilon_{i}=\frac{y+u}{\gamma^{2}},\epsilon_{f}=\frac{y}{\gamma^{2}}\right) dy
\end{equation}
as defined by, e.g., KL1961. The energy emitted per unit time and unit volume is then given by
\begin{equation}
\frac{d\Lambda_{_{\rm ff}}}{d u}=\frac{8 \pi^{2}}{h c^{3}}\left(\frac{2}{3 m_{_{e}}}\right)^{3/2} e^{6} z^{2}n_{e} 
n_{Z,z}\, k_{B} T \frac{2}{\sqrt{\pi}}\frac{e^{-u}}{(k_{B}T)^{1/2}}\, \langle g_{_{\rm ff}}(\gamma^{2},u)\rangle_{em},
\end{equation}
which can be written as
\begin{equation}
\label{eq7}
\frac{d\Lambda_{_{\rm ff}}}{d u}=C\,z^{2}n_{e} n_{_{Z,z}} T^{1/2} e^{-u}\,\langle g_{_{\rm ff}}(\gamma^{2},u)\rangle_{em},
\end{equation}
with \[\displaystyle C= 16 \left(\frac{2 \pi}{3 m_{_{e}}}\right)^{3/2} \frac{e^{6} k_{B}^{1/2}}{hc^{3}} 
=1.4256\times 10^{-27}\mbox{~~erg\,cm$^{3}$\,s$^{-1}$\,K$^{-1/2}$}.\]

\subsection{$n-$distribution of electrons}
For electrons with a $n$-distribution (see Appendix A) 
\begin{equation}
f(E)dE=\frac{2}{\sqrt{\pi}(k_{B}T)^{3/2}}B_{n} E^{1/2}\left(\frac{E}{k_{B}T}\right)^{(n-1)/2}e^{-E/k_{B}T}dE
\end{equation}
where $\displaystyle B_{n}=\frac{\sqrt{\pi}}{2\Gamma(n/2+1)}$, the thermally averaged Gaunt factor given by (\ref{new1}) is
\begin{equation}
\langle g_{_{\rm ff}}(\gamma^{2},u)\rangle_{em}=\frac{1}{N_{em}}\frac{2 e^{-u}}{\sqrt{\pi} (k_{B}T)^{1/2}} B_{n} \int_{0}^{+\infty} 
(y+u)^{(n-1)/2} e^{-y} g_{_{\rm ff}}\left(\epsilon_{i}=\frac{y+u}{\gamma^{2}},\epsilon_{f}=\frac{y}{\gamma^{2}}\right) dy.
\end{equation}
Noting that the incomplete Gamma function \citep{olver2010}
\begin{equation}
\Gamma\left(\frac{n+1}{2},u\right)=e^{-u}\int_{0}^{+\infty} (y+u)^{(n-1)/2} e^{-y} dy,
\end{equation}
the boundary condition for the thermally averaged Gaunt factor implies that 
\begin{equation}
N_{em}=\frac{2}{\sqrt{\pi}}\frac{B_{n}}{(k_{B}T)^{1/2}}\Gamma\left(\frac{n+1}{2},u\right)
\end{equation}
and
\begin{equation}
\label{eq12}
\langle g_{_{\rm ff}}(\gamma^{2},u)\rangle_{em}=\frac{1}{e^{u}\Gamma\left(\frac{n+1}{2},u\right)}\int_{0}^{+\infty} 
(y+u)^{(n-1)/2} e^{-y} g_{_{\rm ff}}\left(\epsilon_{i}=\frac{y+u}{\gamma^{2}},\epsilon_{f}=\frac{y}{\gamma^{2}}\right) dy.
\end{equation}
The amount of energy emitted per unit time and unit volume is then given by
\begin{equation}
\label{eq13}
\displaystyle \frac{d\Lambda_{_{\rm ff}}}{du}= C\,z^{2}n_{e} n_{_{Z,z}} T^{1/2}\,B_{n} \,\Gamma\left(\frac{n+1}{2},u\right) \,\langle 
g_{_{\rm ff}}(\gamma^{2},u)\rangle_{em}.
\end{equation}
\subsection{Total power}
Integration of (\ref{eq7}) and (\ref{eq13}) over the photon frequency spectrum gives the total free-free power associated to an 
ion $(Z,z)$
\begin{equation}
\label{eq6}
\Lambda_{_{\rm ff}}(T)=C\, z^{2} n_{e}n_{_{Z,z}}T^{1/2} \langle g_{_{\rm ff}}(\gamma^{2})\rangle \mbox{~~erg cm$^{-3}$ s$^{-1}$}
\end{equation}
where $\langle g_{_{\rm ff}}(\gamma^{2})\rangle $ is the total free-free Gaunt factor given by
\begin{equation}
\label{eq15}
\langle g_{_{\rm ff}}(\gamma^{2})\rangle = \int_{0}^{+\infty} e^{-u}\, \langle  g_{_{\rm ff}}(\gamma^{2},u) \rangle_{em}\, du.
\end{equation}
for the Maxwell-Boltzmann distribution and
\begin{equation}
\label{eq16}
\langle g_{_{\rm ff}}(\gamma^{2})\rangle =B_{n} \int_{0}^{+\infty}\langle g_{_{\rm ff}}(\gamma^{2},u) \rangle_{em}\,  
\Gamma\left(\frac{n+1}{2},u\right) du.
\end{equation}
for the $n-$distribution of electrons.

From (\ref{eq12}) and (\ref{eq16}) with $n=1$ and noting that $\Gamma\left(1,u\right)=e^{-u}$ and $B_{1}=1$ both the thermally 
averaged and the total free-free Gaunt factors for a Maxwellian distribution of electrons are recovered.

\subsection{Photo emission vs. photo absorption}
Similarly to the definition of the thermally averaged Gaunt factors for photo emission we can define the thermally averaged Gaunt 
factor for photo absorption as
\begin{equation}
\langle g_{_{\rm ff}}(\gamma^{2},u) \rangle_{a} =\frac{1}{N_{a}}\displaystyle \int_{0}^{+\infty}\frac{1}{E^{1/2}}f(E) \, g_{_{\rm 
		ff}}\left(\epsilon_{i}=\frac{E}{z^{2} \mbox{Ry}},\epsilon_{f}=\frac{E+h\nu}{z^2\mbox{Ry}}\right) dE
\end{equation}
with $N_{a}$ being the normalization coefficient determined by the boundary condition $\langle g_{_{\rm ff}}(\gamma^{2},u) 
\rangle_{a} =1$ when $g_{_{\rm ff}}=1$. Using the change of variables $x=E/k_{B}T$ and after determining the normalization 
coefficient, the thermally averaged Gaunt factor for photo absorption is given by 
\begin{equation}
\label{abs1}
\langle g_{_{\rm ff}}(\gamma^{2},u) \rangle_{a} =\int_{0}^{+\infty} e^{-x} g_{_{\rm 
		ff}}\left(\epsilon_{i}=\frac{x}{\gamma^{2}},\epsilon_{f}=\frac{x+u}{\gamma^{2}}\right) dx
\end{equation}
for the Maxwell-Boltzmann distribution and
\begin{equation}
\label{abs2}
\langle g_{_{\rm ff}}(\gamma^{2},u) \rangle_{a} =\frac{1}{\Gamma\left(\frac{n+1}{2}\right)}\int_{0}^{+\infty} x^{(n-1)/2} e^{-x} 
g_{_{\rm ff}}\left(\epsilon_{i}=\frac{x}{\gamma^{2}},\epsilon_{f}=\frac{x+u}{\gamma^{2}}\right) dx
\end{equation}
for the $n$-distribution. The Maxwellian temperature averaged Gaunt factors for photo emission and photo absorption, 
equations (\ref{eq5}) and (\ref{abs1}), yield the same result. This is a consequence of the principle of detailed balance that is 
required for thermodynamic equilibrium to be reached \citep[see, e.g.,][]{armstrong1971}. A condition that is not verified for the 
$n-$distribution with $n>1$. Thus,  equations (\ref{eq12}) and (\ref{abs2}) have different results.

\begin{figure*}[thbp]
	\centering
	\includegraphics[width=0.45\hsize,angle=0]{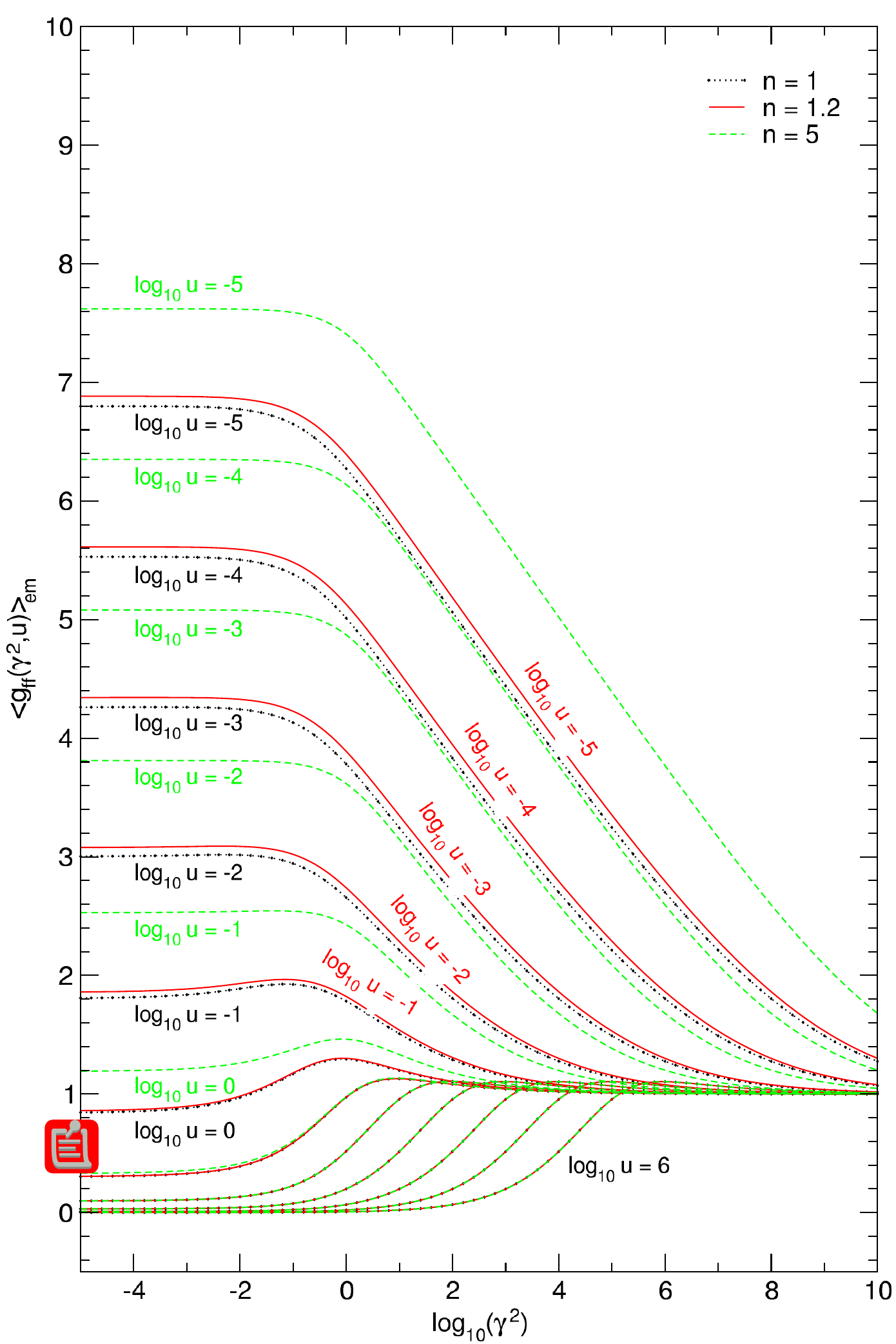}\includegraphics[width=0.45\hsize,angle=0]{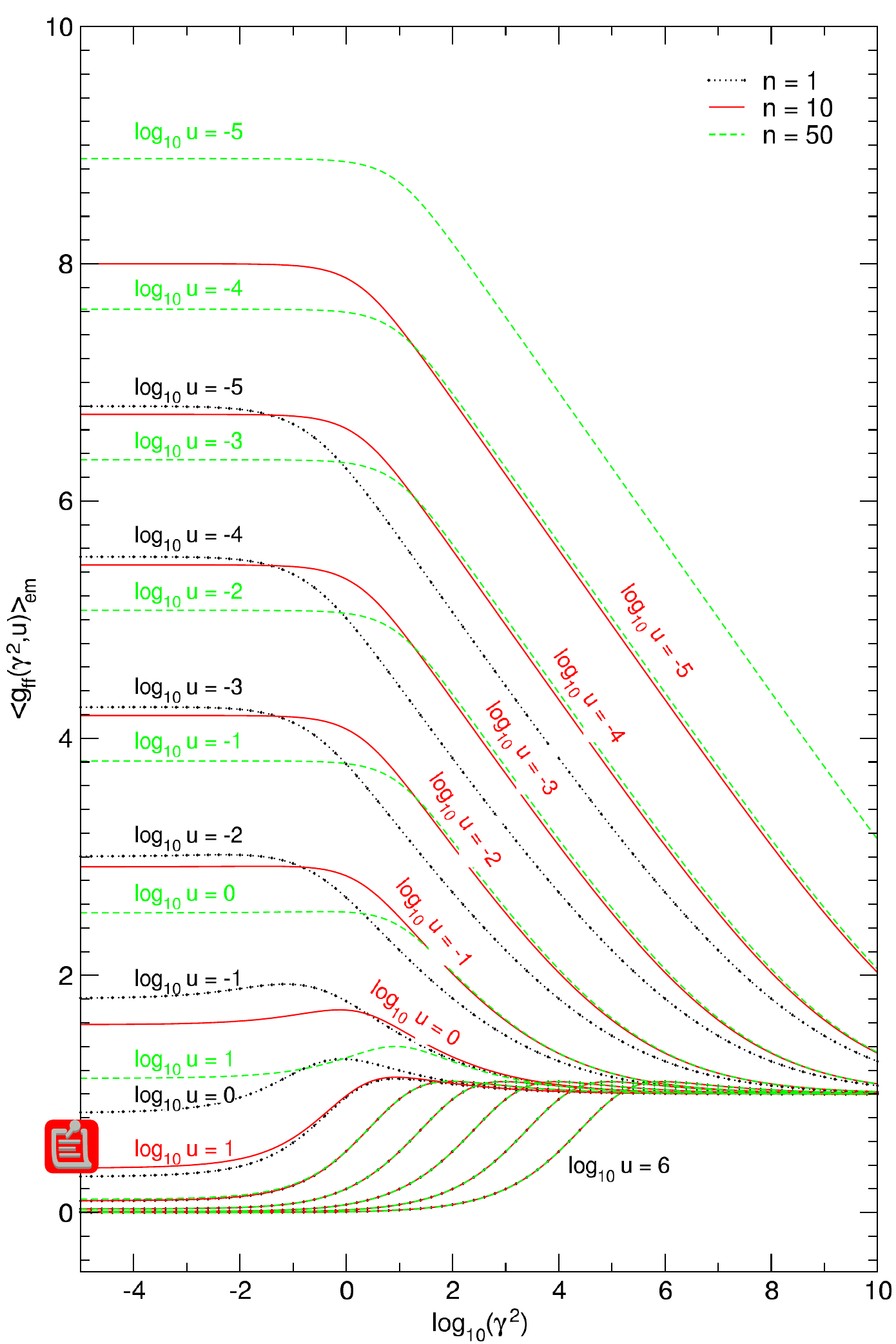}
	\caption{Thermally averaged Gaunt factors for electrons with an $n-$distribution with $n=1$, 1.2, and 5 (left panel), and 
		$n=1$, 10 and 50 (right panel). The curves are displayed for $\log_{10}(u)$ varying from -5 to 6  from top to bottom in 
		each panel.}
	\label{averaged}
\end{figure*}

\begin{figure*}[thbp]
	\centering
	\includegraphics[width=0.45\hsize,angle=0]{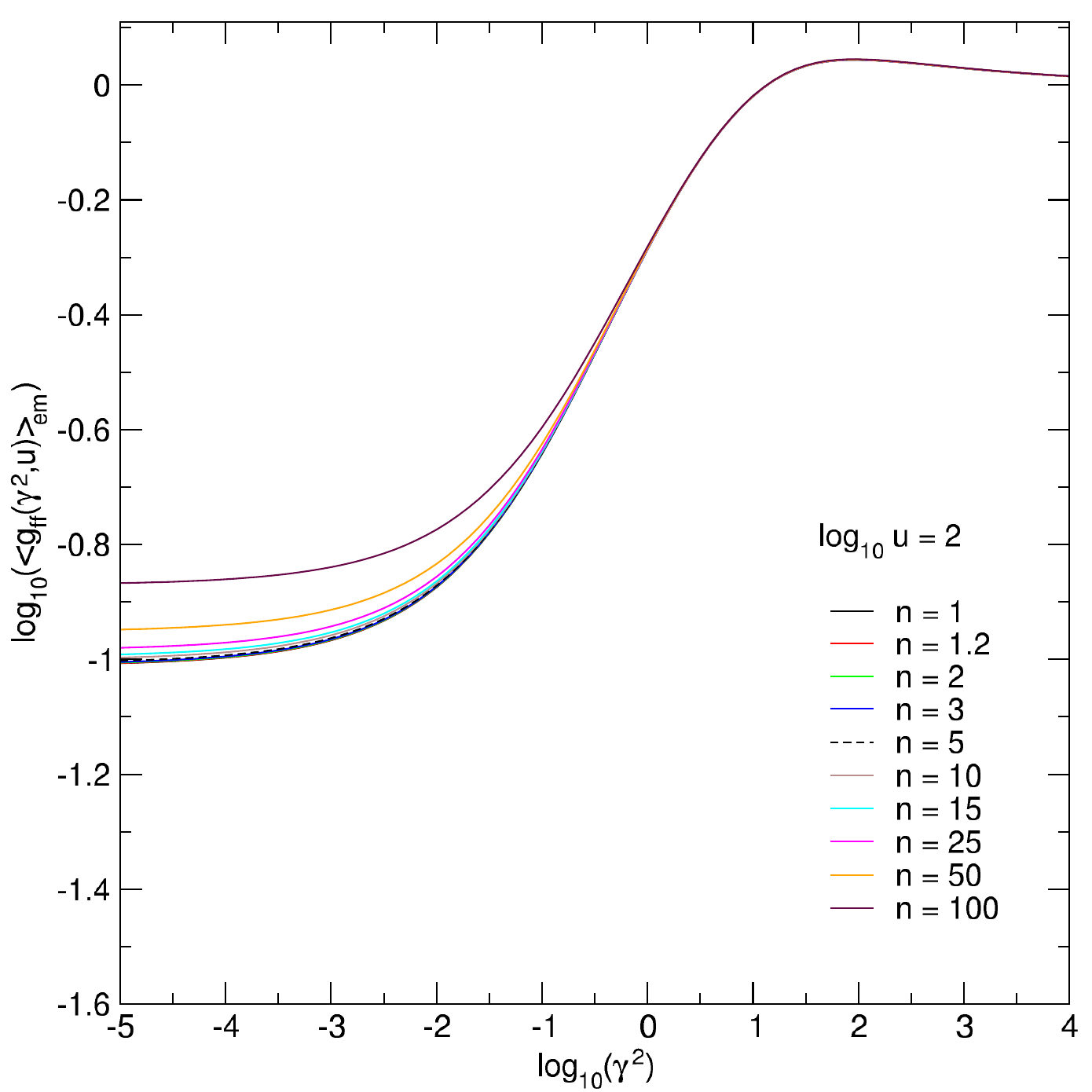}\includegraphics[width=0.45\hsize,angle=0]{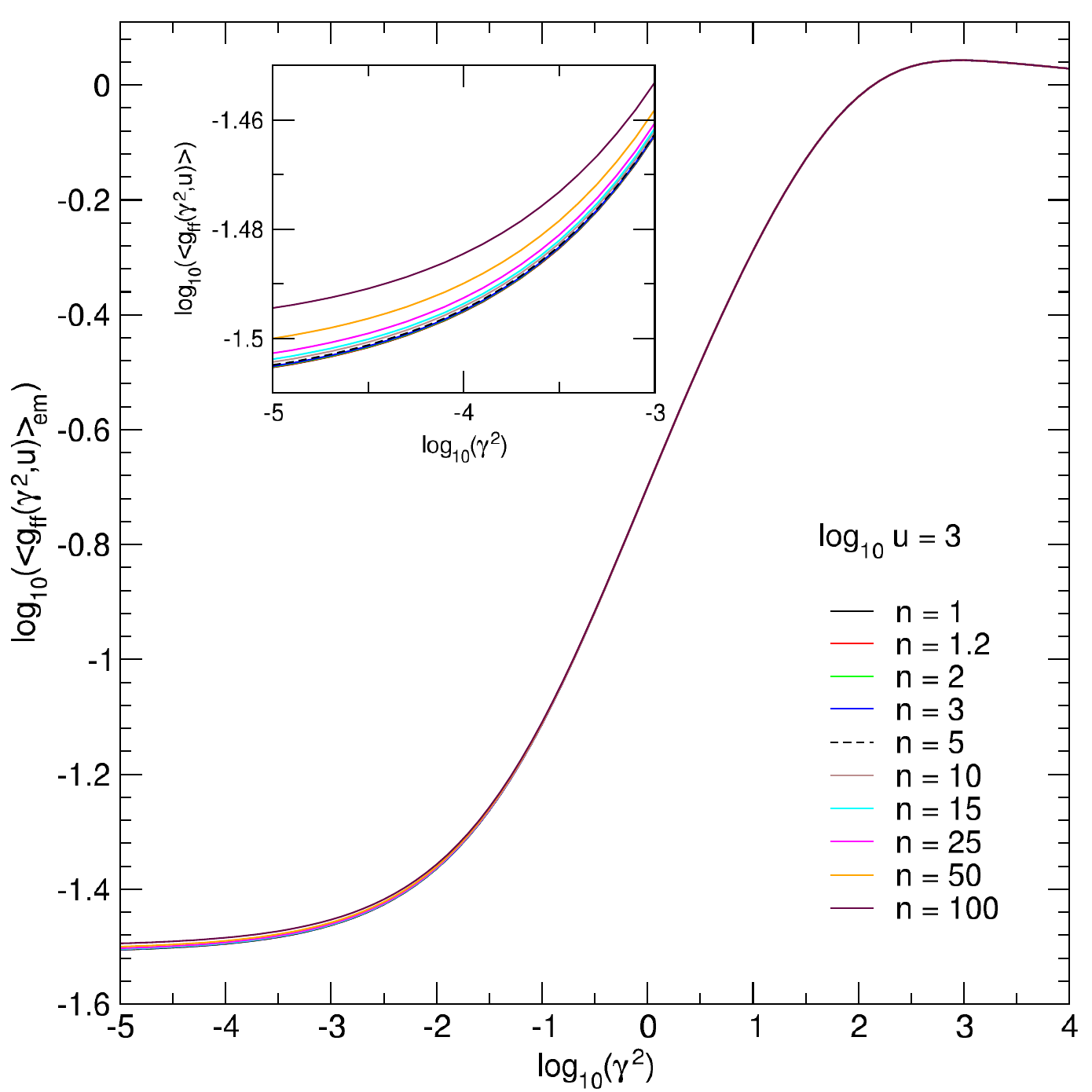}
	\caption{Detail of the thermally averaged Gaunt factors (in logarithmic scale) for electrons with an $n-$distribution ($n$ 
		varying from 1 to 100) as function of $\gamma^{2}$ for $\log_{10}u=2$ and 3. Although it appears from the previous 
		figure that $\langle g_{_{\rm ff}}(\gamma^{2},u)\rangle_{em}$ overlap for $\log_{10}u>1$ there are still small 
		differences at $\log_{10}u=3$ for $\gamma^{2}<0$ (inset plot in the right panel).}
	\label{averaged_details}
\end{figure*}

\section{Calculations and Results}

We calculated numerically the temperature averaged, $\langle g_{_{\rm ff}}(\gamma^{2},u) \rangle_{em}$, and total, $\langle 
g_{_{\rm ff}}(\gamma^{2})\rangle$, free-free Gaunt factors for an $n-$distribution of electrons ($n$ ranging from 1 
(Maxwell-Boltzmann distribution) to 100) for $\gamma^{2} \in [10^{-5},10^{10}]$ and $u\in[10^{-12},10^{11}]$ using 
the Gaunt factors associated to free-free absorption by an electron in the presence of a Coulomb field of an 
ion. For completeness the range in $\gamma^{2}$ is limited to $10^{-5}$ corresponding to an upper temperature of
$T_{e}/Z^{2}=1.578\times 10^{10}$ K, although at this temperature relativistic effects are seen and electron-electron 
bremsstrahlung dominates \citep[see, e.g.,][]{itoh1985}.

The calculations proceeded as follows: first the free-free Gaunt factors for the absorption of photons of frequency $h\nu$ 
by electrons with a range of possible initial energies $E_{i}$ were explicitly calculated (see AB2015), followed by the 
determination of the temperature averaged and total Gaunt factors (equations \ref{eq5} and \ref{eq7}, respectively). The Gaunt 
factors calculation involved double and quad-precisions (see discussion in AB2015), while the numerical integrations were 
performed with a precision of $10^{-15}$ using the double-exponential over a semi-finite interval method of \citet{takahasi1974} 
and \citet{mori2001}. We use a modified version (parallelized version) of the Numerical Automatic Integrator for 
Improper Integral package developed by T. Ooura\footnote{http://www.kurims.kyoto-u.ac.jp/\textasciitilde ooura/intde.html}.

Figure~\ref{averaged} displays the temperature averaged Gaunt factors calculated for electrons having an $n-$distribution with 
$n=1$ (displayed in the two panels by black dotted lines), 1.2, 5, 10 and 50. The curves are displayed for $\log_{10}(u)$ varying 
from -5 to 6  from top to bottom in each panel, although our calculations consider a wider range in $n$, $\gamma^{2}$ and $u$ 
(see Section 5). Figure~\ref{averaged_details} details $\langle g_{_{\rm ff}}(\gamma^{2},u) \rangle_{em}$ as function of 
$\gamma^{2}$ for $\log_{10}u=2$ (left panel) and 3 (right panel) in logarithmic scale to show the deviations of $\langle g_{_{\rm  
ff}}(\gamma^{2},u) \rangle_{em}$ for different $n$ at these frequencies.

The temperature averaged Gaunt factors for $n>1$ deviates from those calculated for the Maxwell-Boltzmann distribution 
($n=1$). As $n$ increases the deviations increase with the decrease of $\log_{10}u$ (Figure~\ref{averaged}). For $\log_{10}u>1$ 
$\langle g_{_{\rm ff}}(\gamma^{2},u) \rangle_{em}$ converges to the Maxwell-Boltzmann value (Figure~\ref{averaged}) overlapping 
it when $\log_{10}u>3$. There are still small differences at $\log_{10}u=3$ for $\gamma^{2}<0$ (Figure~\ref{averaged_details}).

\begin{figure}[thbp]
	\centering
	\includegraphics[width=0.45\hsize,angle=0]{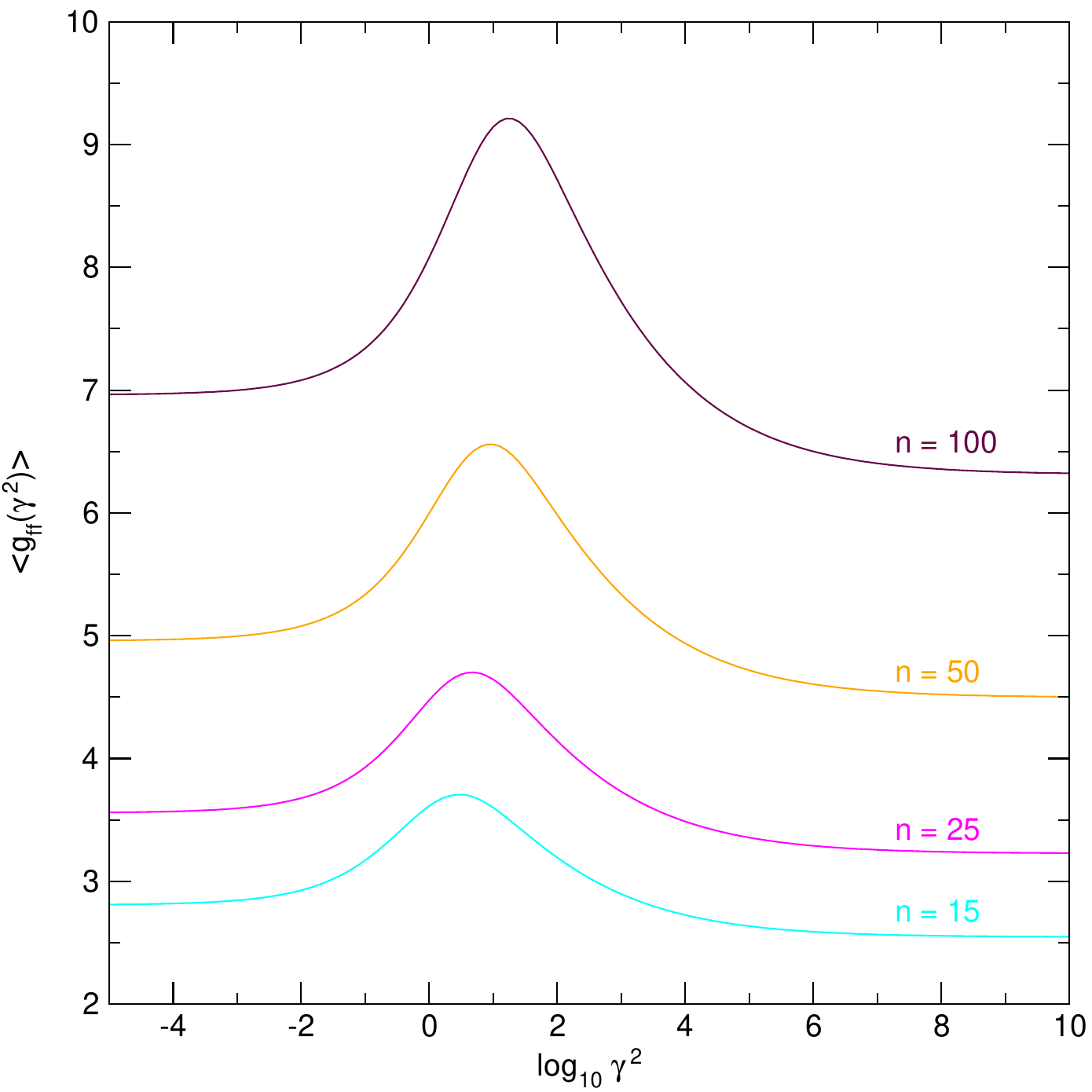}\\
	\includegraphics[width=0.45\hsize,angle=0]{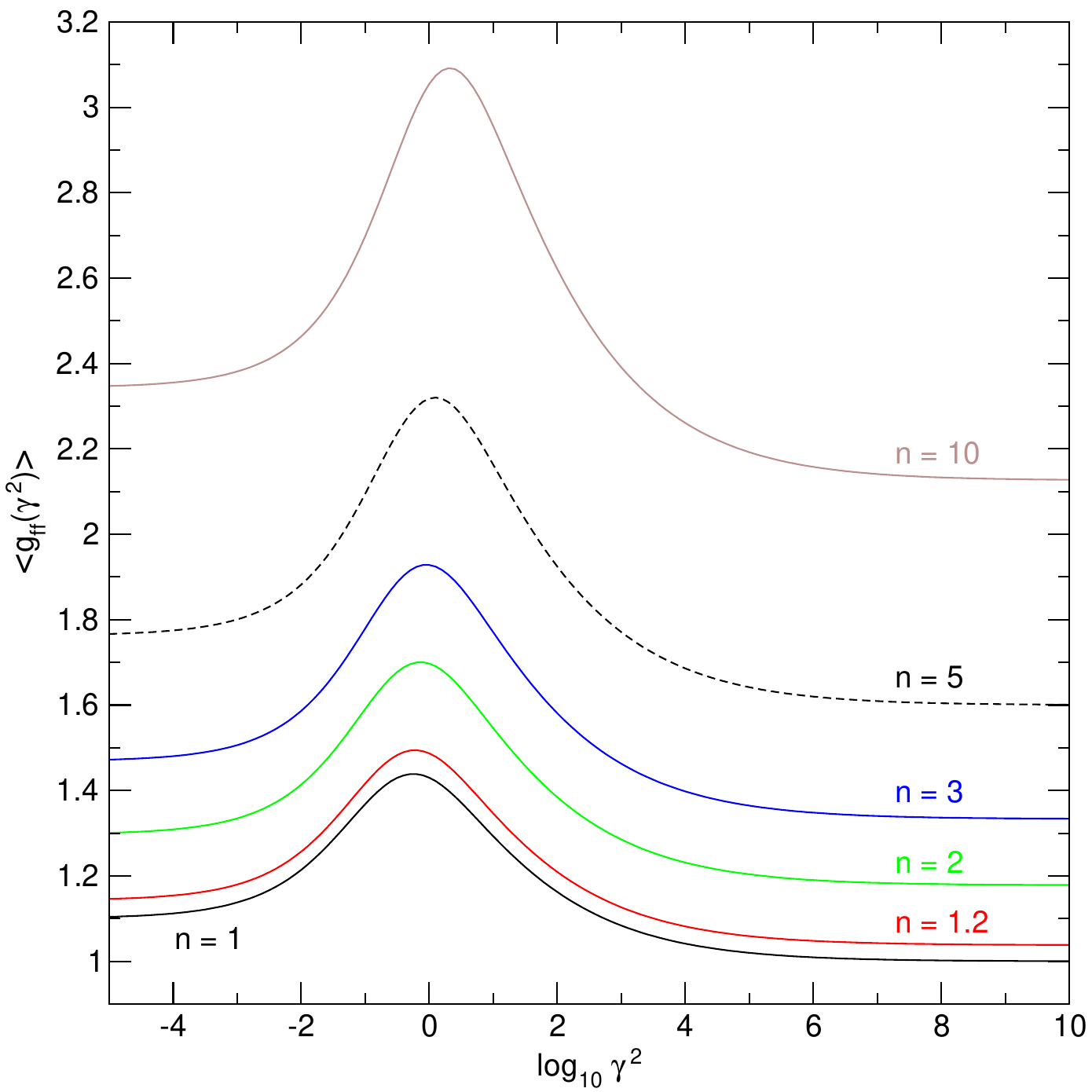}
	\caption{Total free-free Gaunt factor calculated for the $n-$distributed electrons with $n=1$, 1.2, 2, 3, 5, and 10 (bottom 
		panel) and $n=15$,  25, 50, and 100 (top panel). With increasing $n$, the total Gaunt increases and its peak maximum 
		moves to the right. \label{total_ff}}
\end{figure}

Figure~\ref{total_ff} displays the total Gaunt factor, $\langle g_{_{\rm ff}}(\gamma^{2}) \rangle$, variation with $\gamma^{2}$ 
calculated for the $n=1$, 1.2, 2, 3, 5 and 10 (bottom panel) and 15, 25, 50 and 100 (top panel). For each $n$ the total Gaunt 
factor has a similar profile as in $n=1$ but with the peak shifted to the right. In addition, and as expected from the thermally 
averaged Gaunt factors evolution with $\gamma^{2}$, $\langle g_{_{\rm ff}} (\gamma^{2})\rangle$ increases with increasing $n$.

\section{Validation of the Maxwellian thermally averaged and total Gaunt factors}

\begin{figure}[thbp]
	\centering
	\includegraphics[width=0.45\hsize,angle=0]{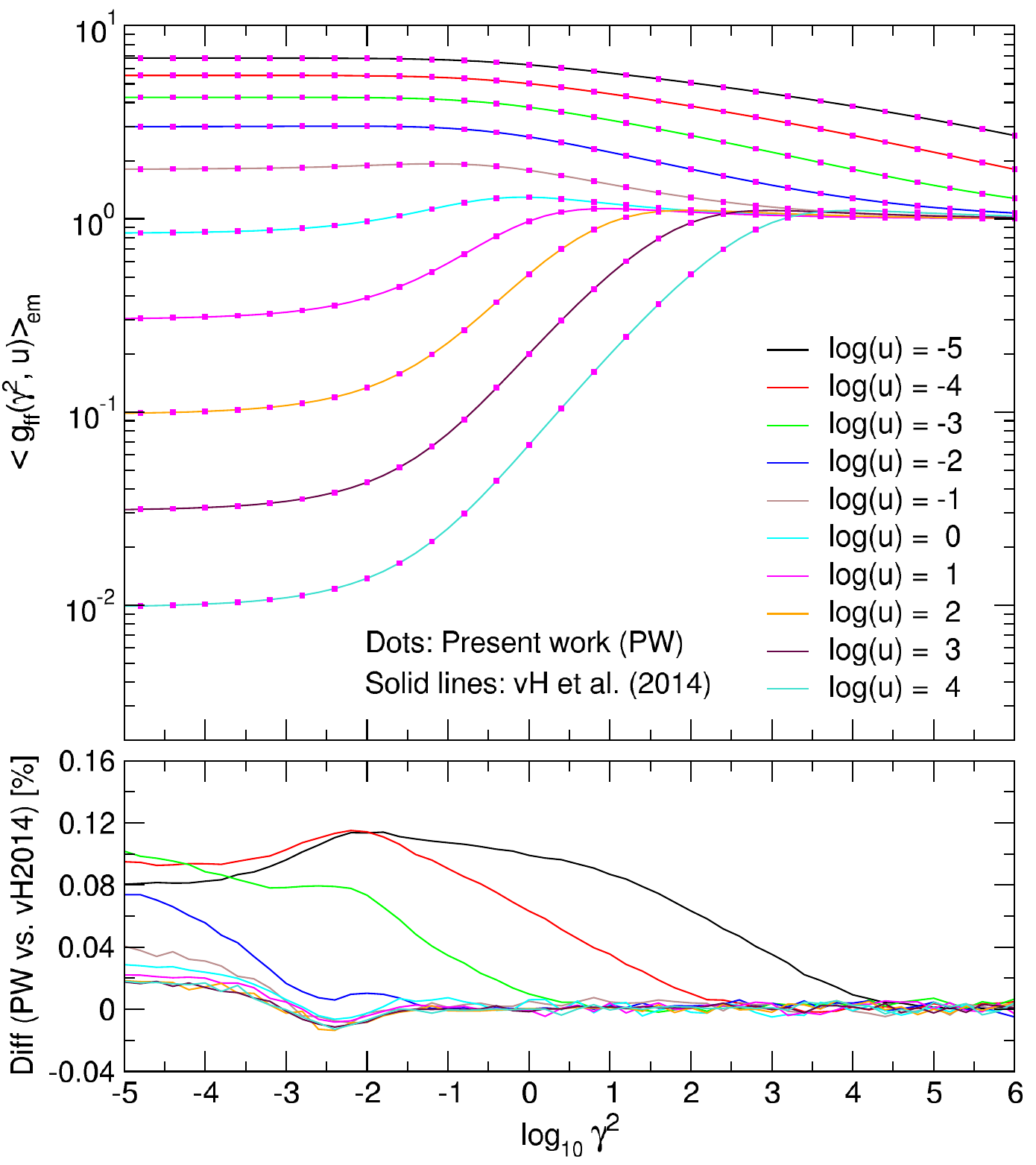}
	\caption{{\rm Top panel:} Comparison between the temperature averaged free-free Gaunt factors for a thermal distribution of 
		electrons published by \citet[][solid lines]{vanhoof2014} and calculated in the present work (magenta dots). 
		{\rm Bottom panel:} Relative differences between the temperature averaged Gaunt factors of obtained in the present work 
		and those of \citet{vanhoof2014}. \label{comparison1}}
\end{figure}
\begin{figure}[thbp]
	\centering
	\includegraphics[width=0.45\hsize,angle=0]{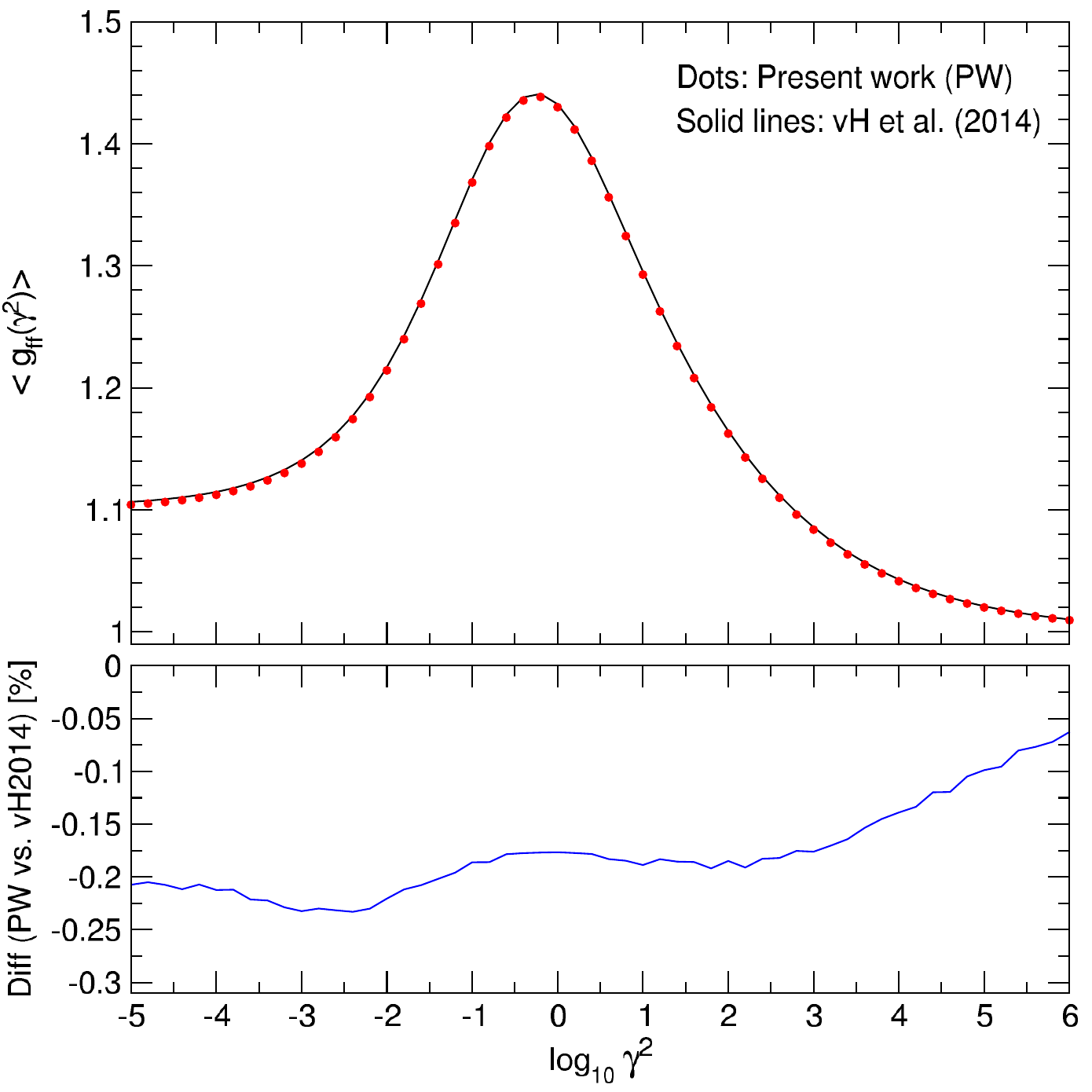}
	\caption{{\rm Top panel:} Comparison of the total free-free Gaunt factors for a thermal distribution of electrons calculated in 
	the present work (red dots) and by van Hoof et al. (2014; black line). {\rm Bottom panel:} Relative difference between the total 
		Gaunt factor obtained in the present work and by \citet{vanhoof2014}.
		\label{comparison2}}
\end{figure}

In order to validate our results, we carried out a detailed comparison of the temperature averaged and total free-free Gaunt 
factors, calculated for electrons with a Maxwellian temperature, with those published by \citet{vanhoof2014}. The top panels of 
the Figures~\ref{comparison1} and \ref{comparison2} compare the temperature averaged and total free-free  Gaunt factors 
determined for the Maxwell-Boltzmann distribution in the present work (red dots) and those calculated by \citet[][solid 
lines]{vanhoof2014}. 

The bottom panel of Figure~\ref{comparison1} displays the relative difference between the two calculations of the temperature 
averaged Gaunt factor in the ranges $-5 \leq \log(\gamma^{2})\leq 6$ and $-5 \leq \log(u)\leq 4$. At first sight there seems to be 
no variation between the solid lines and the dots overlaying them. In fact the two calculations have relative differences (in 
percentage) smaller than 0.12\% for $\log(u)=-5$ and -4 and smaller than 0.04\% for $\log(u)\geq -1$. The relative difference 
between the total Gaunt factors obtained in the two calculations is smaller than 0.233\% (bottom panel of 
Figure~\ref{comparison2}).

These relative differences result from the different numerical techniques to calculate the hypergeometric functions and 
machine precision adopted for the calculations of the Gaunt factors (see discussions in the above mentioned papers and in 
AB2015), as well as the adopted integrating method used in the calculations of the thermally averaged and total Gaunt factors. 
While in the present work the double-exponential quadrature method extended to any real function \citep{mori2001} was 
used to calculate the different integrals, van Hoof et al. used an adaptive step-size algorithm for the integrations with an 
estimate of the  remainder of the integral to infinity being smaller than $10^{-7}$ times the total integral up to that point. The 
relative differences shown above are indicative of the adequacy of our results.

\section{Tables}

In Appendices B and C, a set of Tables 1-11 referring to the temperature averaged  ($\langle g_{_{\rm ff}}(\gamma^{2},u)\rangle_{em}$ 
vs. $\gamma^{2}$ for different $u$) and the total ($g_{_{\rm ff}}(\gamma^{2})$ vs. $\gamma^{2}$) free-free Gaunt factor are 
presented. The parameter space comprises $1\leq n\leq 100$, $\gamma^{2}\in[10^{-5},10^{10}]$, and $u\in [10^{-5},10^{6}]$. Extended data 
with $u\in [10^{-12},10^{11}]$ are included as supplementary material with this paper or can be requested from the 
authors.

\section{Application to an optically thin plasma}

We calculated the electron-ion\footnote{Note that electron-electron bremsstrahlung, although important at $10^{9}$ K, is not included in the 
present calculation.} free-free contribution to the radiative losses of a plasma composed of H, He, C, N, O, Ne, Mg, Si, S, and Fe and having 
solar abundances \citep{asplund2009} by considering the evolution of a gas parcel freely cooling from $10^{9}$ K and evolving under 
collisional ionization equilibrium (CIE). Besides the free-free emission, the other physical processes included in this calculation comprise 
electron impact ionization, inner-shell excitation, auto-ionization, radiative and dielectronic recombination. The internal energy of the gas 
parcel includes the contributions due to the thermal translational energy plus the energy stored in ionization.

The atomic data used in the present calculations includes the electron impact ionization cross sections discussed in \citet{dere2007} and 
available in the Chianti database \citep[see, e.g.,][]{dere2009}. From these cross sections we calculate the ionization rates associated to an 
ion of atomic number $Z$ and ionic charge $z$ by averaging the product $\sigma(E) v$ over the impacting particle kinetic energy distribution 
$f(E)$
\begin{equation}
\label{eq_rate}
\langle \sigma v\rangle =\int_{I_{_{Z,z}}}^{+\infty} \sigma(E) (2E/m_{e})^{1/2} f(E) dE\mbox{~~cm$^{3}$ s$^{-1}$}
\end{equation}
where $m_{e}$ is the electron mass, and $I_{_{Z,z}}$ is the threshold energy in eV. 

The radiative and dielectronic recombination rates for the $n$ distribution are calculated from the fit coefficients to the Maxwellian rates 
\citep{dzifcakova1998}, that is, the radiative recombination rates for the $n$-distribution are determined from
\begin{equation}
\alpha_{RR}^{n}=\alpha_{_{Z,z}}^{MB}B_{n}\frac{\Gamma(n/2-\eta+1)}{\Gamma(3/2-\eta)}
\end{equation}
where $\alpha_{RR}^{MB}$ is the Maxwellian radiative recombination rate, and $\eta$ is a parameter of the power-law fit of the 
Maxwellian rate \citep[see, e.g.,][]{woods1981}
\begin{equation}
\label{power_law}
\alpha_{RR}^{MB}=A_{rad}\left(\frac{T}{10^{4}\mbox{K}}\right)^{-\eta}\mbox{~~cm$^{3}$ s$^{-1}$}.
\end{equation}
The dielectronic recombination rates for the $n$-distribution of electrons are determined from coefficients of the 
Maxwellian rates given by the \citet{burgess1965} general formula
\begin{equation}
\alpha_{DR}^{MB}=\frac{1}{(k_{B}T)^{3/2}}\sum_{j}c_{j} e^{-E_{j}/(k_{B}T)} \mbox{~~cm$^{3}$ s$^{-1}$}
\end{equation}
through \citep{dzifcakova1998}
\begin{equation}
\label{dr_rate}
\alpha_{DR}^{n}=\frac{B_{n}}{(k_{B}T)^{3/2}}\sum_{j}c_{j} 
\left(\frac{E_{j}}{k_{B}T}\right)^{(n-1)/2}e^{-E_{j}/(k_{B}T)}\mbox{~~cm$^{3}$ s$^{-1}$.}
\end{equation}
The Maxwellian radiative recombination rate coefficients are taken from 
\citet{badnell2006ApJS}\footnote{amdpp.phys.strath.ac.uk/tamoc/DATA/} for all bare nuclei through Na-like ions recombining to H through 
Mg-like ions, \citet{altun2007} for Mg-like ions, \citet{abdel-naby2012} for Al-like ions, \citet{nikolic2010} for Ar-like ions, and 
\cite{badnell2006ApJ} for \fexiii-\fex ions. The Maxwellian dielectronic recombination rates are taken from \citet{badnell2006A&A} for H-like 
ions, \citet{bautista2007} for He-like ions, \citet{colgan2004,colgan2003} for Li and Be-like ions, \citet{altun2004,altun2006,altun2007} for 
B\footnote{For \nevi and \mgvii we use the erratum; \citet{altun2005}}, Na and Mg-like ions, \citet{zat2003,zat2004a,zat2004b,zat2006} for 
C\footnote{For \nitii we follow the erratum; \citet{zat2005b}}, O\footnote{For \neiii, \mgv, \fexix we use the erratum; \citet{zat2005a}}, F and 
Ne-like ions, \citet{mitnik2004} for N-like ions, \citet{abdel-naby2012} for Al-like ions, and \citet{nikolic2010} for Ar-like ions. Radiative and 
dielectronic recombination rates for \sulii, \suliii and \fevii are adopted from \citet{mazzotta1998A&AS..133..403M}, while for the remaining ions 
we adopt the radiative and dielectronic recombination rates derived with the unified electron-ion recombination method 
\citep{nahar1994PhRvA..49.1816N} and available at NORAD-Atomic-Data\footnote{www.astronomy.ohio-state.edu/$\sim$nahar}.

The calculation was carried out with the Collisional + Photo Ionization Plasma Emission Software (CPIPES; de Avillez 2017 in 
preparation), which is a complete rewrite of the E+AMPEC code \citep{avillez2012ASPC..453..341D}. The ionization structure due 
to the 102 ions and 10 atoms (in a total of 112 linear equations) of the gas parcel evolving under CIE conditions were calculated at 
each temperature using a Gauss elimination method with scaled partial pivoting \citep{cheney2008} and a tolerance of 
$10^{-15}$. Having calculated the ionization structure of the gas parcel at each temperature, the losses of energy due to 
free-free emission are calculated using equation (7) and assuming the hydrogenic approximation.

\begin{figure}[thbp]
	\centering
	\includegraphics[width=0.45\hsize,angle=0]{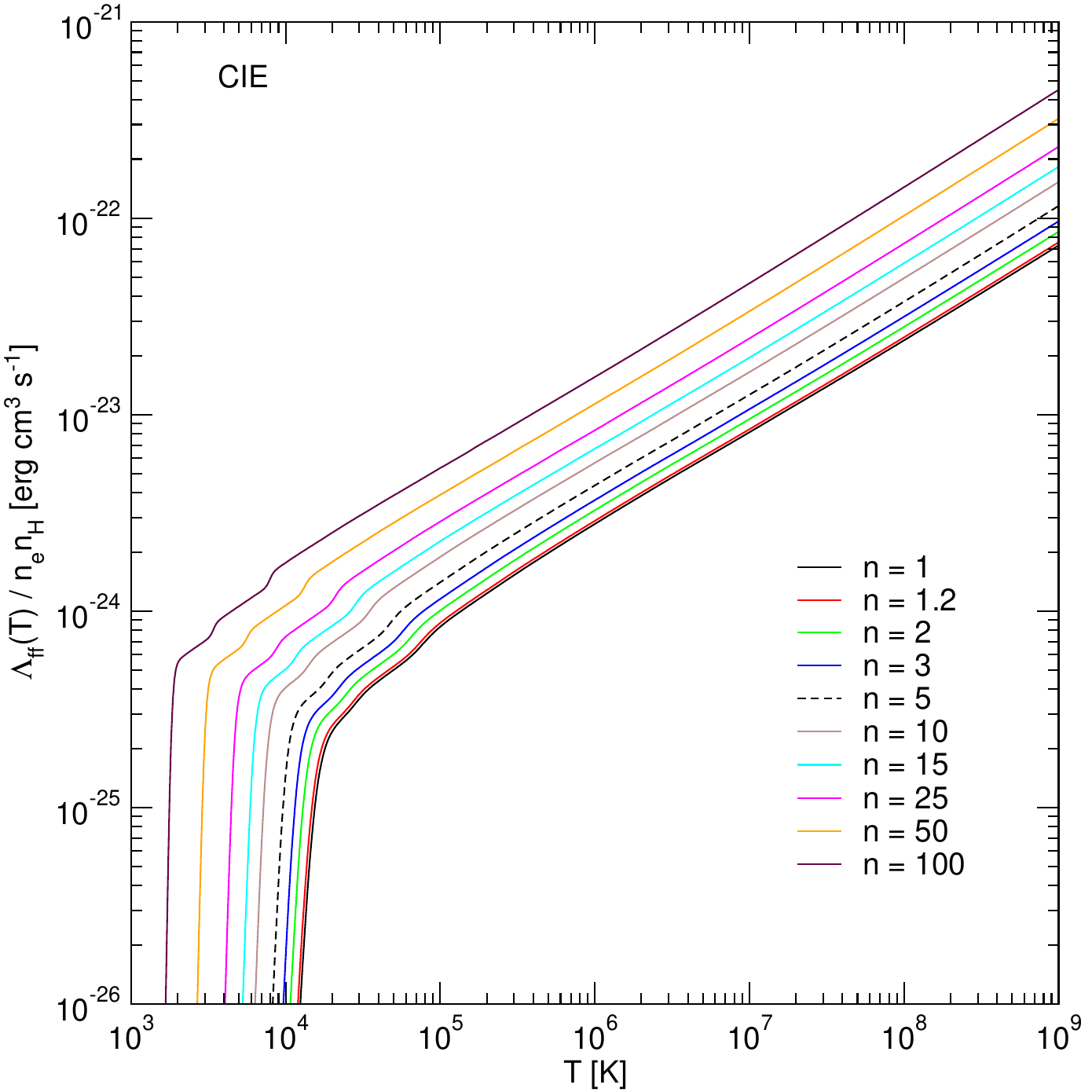}
	\caption{Normalized (to $n_{e}n_{H}$) radiative losses due to free-free  emission by an optically thin plasma evolving under 
		collisional ionization equilibrium and for electrons having a $n-$distribution ($n=1$, 1.2, 2, 3, 5, 10, 15, 25, 50, and 100).}
	\label{cie_cooling}
\end{figure}

Figure~\ref{cie_cooling} displays the free-free emission resulting from the CIE calculation for electrons described by the 
$n-$distribution with $n$ ranging from 1 (Maxwell-Boltzmann distribution) through 100. The free-free emission increases 
with $n$ as expected from the variation of the total Gaunt factor with $n$ as shown in Figure~\ref{total_ff}. This loss of 
radiation has its minimum when electrons relax to thermal equilibrium, that is, when $n$ becomes 1. 

\section{Final remarks}

We carried out a detailed calculation of the temperature averaged and total free-free Gaunt factors that are useful for the 
determination of the free-free emission power spectrum and losses of radiation in astrophysical plasmas. Bremsstrahlung is in 
many situations the most important electromagnetic information we can obtain, e.g. in clusters of galaxies with plasma 
temperatures in the keV range. It is well-known that galaxy cluster or group mergers can generate large scale shock waves \citep[see, 
e.g.,][]{russell2014} and/or high temperature gradients at the interface of different intracluster gases, both of which would 
drive the electron population, responsible for the emission, out of equilibrium. It is therefore desirable to provide tables of 
free-free Gaunt factors over a wide range in temperatures for a variety of $n-$distributions.  

\acknowledgements
We acknowledge the useful comments by the anonymous referee. This research was supported by the project "Hybrid computing using 
accelerators \& coprocessors-modelling nature with a novell approach" (PI: M.A.) funded by the InAlentejo program, CCDRA, Portugal. Partial 
support to M.A. and D.B. was provided by the \emph{Deut\-scheFor\-schungs\-ge\-mein\-schaft}, DFG project ISM-SPP 1573. The 
computations were carried out with the ISM/SKA Xeon Phi supercomputer of the Computational Astrophysics Group, University of \'Evora.

\bibliography{bibliography}

\begin{thebibliography}{}
\expandafter\ifx\csname natexlab\endcsname\relax\def\natexlab#1{#1}\fi
\providecommand{\url}[1]{\href{#1}{#1}}

\bibitem[{{Abdel-Naby} {et~al.}(2012){Abdel-Naby}, {Nikoli{\'c}}, {Gorczyca},
  {Korista}, \& {Badnell}}]{abdel-naby2012}
{Abdel-Naby}, S.~A., {Nikoli{\'c}}, D., {Gorczyca}, T.~W., {Korista}, K.~T., \&
  {Badnell}, N.~R. 2012, \aap, 537, A40

\bibitem[{{Altun} {et~al.}(2004){Altun}, {Yumak}, {Badnell}, {Colgan}, \&
  {Pindzola}}]{altun2004}
{Altun}, Z., {Yumak}, A., {Badnell}, N.~R., {Colgan}, J., \& {Pindzola}, M.~S.
  2004, \aap, 420, 775

\bibitem[{{Altun} {et~al.}(2005){Altun}, {Yumak}, {Badnell}, {Colgan}, \&
  {Pindzola}}]{altun2005}
---. 2005, \aap, 433, 395

\bibitem[{{Altun} {et~al.}(2006){Altun}, {Yumak}, {Badnell}, {Loch}, \&
  {Pindzola}}]{altun2006}
{Altun}, Z., {Yumak}, A., {Badnell}, N.~R., {Loch}, S.~D., \& {Pindzola}, M.~S.
  2006, \aap, 447, 1165

\bibitem[{{Altun} {et~al.}(2007){Altun}, {Yumak}, {Yavuz}, {Badnell}, {Loch},
  \& {Pindzola}}]{altun2007}
{Altun}, Z., {Yumak}, A., {Yavuz}, I., {et~al.} 2007, \aap, 474, 1051

\bibitem[{{Armstrong}(1971)}]{armstrong1971}
{Armstrong}, B.~H. 1971, \jqsrt, 11, 1731

\bibitem[{{Asplund} {et~al.}(2009){Asplund}, {Grevesse}, {Sauval}, \&
  {Scott}}]{asplund2009}
{Asplund}, M., {Grevesse}, N., {Sauval}, A.~J., \& {Scott}, P. 2009, \araa, 47,
  481

\bibitem[{{Badnell}(2006{\natexlab{a}})}]{badnell2006ApJS}
{Badnell}, N.~R. 2006{\natexlab{a}}, \apjs, 167, 334

\bibitem[{{Badnell}(2006{\natexlab{b}})}]{badnell2006ApJ}
---. 2006{\natexlab{b}}, \apjl, 651, L73

\bibitem[{{Badnell}(2006{\natexlab{c}})}]{badnell2006A&A}
---. 2006{\natexlab{c}}, \aap, 447, 389

\bibitem[{{Bautista} \& {Badnell}(2007)}]{bautista2007}
{Bautista}, M.~A., \& {Badnell}, N.~R. 2007, \aap, 466, 755

\bibitem[{{Behringer} \& {Fantz}(1994)}]{behringer1994}
{Behringer}, K., \& {Fantz}, U. 1994, Journal of Physics D Applied Physics, 27,
  2128

\bibitem[{{Berezhko} \& {Ellison}(1999)}]{berezhko1999}
{Berezhko}, E.~G., \& {Ellison}, D.~C. 1999, \apj, 526, 385

\bibitem[{{Buneman}(1963)}]{buneman1963}
{Buneman}, O. 1963, Physical Review Letters, 10, 285

\bibitem[{{Burgess}(1965)}]{burgess1965}
{Burgess}, A. 1965, \apj, 141, 1588

\bibitem[{{Carson}(1988)}]{carson1988}
{Carson}, T.~R. 1988, \aap, 189, 319

\bibitem[{{Cheney} \& {Kincaid}(2008)}]{cheney2008}
{Cheney}, W., \& {Kincaid}, D. 2008, {Numerical Mathematics and Computing}, 6th
  edn. (Thomson Brooks/Cole)

\bibitem[{{Colgan} {et~al.}(2004){Colgan}, {Pindzola}, \&
  {Badnell}}]{colgan2004}
{Colgan}, J., {Pindzola}, M.~S., \& {Badnell}, N.~R. 2004, \aap, 417, 1183

\bibitem[{{Colgan} {et~al.}(2003){Colgan}, {Pindzola}, {Whiteford}, \&
  {Badnell}}]{colgan2003}
{Colgan}, J., {Pindzola}, M.~S., {Whiteford}, A.~D., \& {Badnell}, N.~R. 2003,
  \aap, 412, 597

\bibitem[{{de Avillez} \& {Breitschwerdt}(2015)}]{avillez2015}
{de Avillez}, M.~A., \& {Breitschwerdt}, D. 2015, \aap, 580, A124

\bibitem[{{de Avillez} {et~al.}(2012){de Avillez}, {Spitoni}, \&
  {Breitschwerdt}}]{avillez2012ASPC..453..341D}
{de Avillez}, M.~A., {Spitoni}, E., \& {Breitschwerdt}, D. 2012, in
  Astronomical Society of the Pacific Conference Series, Vol. 453, Advances in
  Computational Astrophysics: Methods, Tools, and Outcome, ed.
  R.~{Capuzzo-Dolcetta}, M.~{Limongi}, \& A.~{Tornamb{\`e}}, 341

\bibitem[{{Dere}(2007)}]{dere2007}
{Dere}, K.~P. 2007, \aap, 466, 771

\bibitem[{{Dere} {et~al.}(2009){Dere}, {Landi}, {Young}, {Del Zanna},
  {Landini}, \& {Mason}}]{dere2009}
{Dere}, K.~P., {Landi}, E., {Young}, P.~R., {et~al.} 2009, \aap, 498, 915

\bibitem[{{Druyvesteyn}(1930)}]{druyvesteyn1930}
{Druyvesteyn}, M.~J. 1930, Zeitschrift fur Physik, 64, 781

\bibitem[{{Dzif{\v c}{\'a}kov{\'a}} {et~al.}(2011){Dzif{\v c}{\'a}kov{\'a}},
  {Homola}, \& {Dud{\'{\i}}k}}]{dzifcakova2011}
{Dzif{\v c}{\'a}kov{\'a}}, E., {Homola}, M., \& {Dud{\'{\i}}k}, J. 2011, \aap,
  531, A111

\bibitem[{{Dzif\v{c}\'akov\'a}(1998)}]{dzifcakova1998}
{Dzif\v{c}\'akov\'a}, E. 1998, \solphys, 178, 317

\bibitem[{{Farley}(1963)}]{farley1963}
{Farley}, D.~T. 1963, Physical Review Letters, 10, 279

\bibitem[{{Hares} {et~al.}(1979){Hares}, {Kilkenny}, {Key}, \&
  {Lunney}}]{hares1979}
{Hares}, J.~D., {Kilkenny}, J.~D., {Key}, M.~H., \& {Lunney}, J.~G. 1979,
  Physical Review Letters, 42, 1216

\bibitem[{{Hummer}(1988)}]{hummer1988}
{Hummer}, D.~G. 1988, \apj, 327, 477

\bibitem[{{Itoh} {et~al.}(1985){Itoh}, {Nakagawa}, \& {Kohyama}}]{itoh1985}
{Itoh}, N., {Nakagawa}, M., \& {Kohyama}, Y. 1985, \apj, 294, 17

\bibitem[{{Itoh} {et~al.}(2000){Itoh}, {Sakamoto}, {Kusano}, {Nozawa}, \&
  {Kohyama}}]{itoh2000}
{Itoh}, N., {Sakamoto}, T., {Kusano}, S., {Nozawa}, S., \& {Kohyama}, Y. 2000,
  \apjs, 128, 125

\bibitem[{{Janicki}(1990)}]{janicki1990}
{Janicki}, C. 1990, Computer Physics Communications, 60, 281

\bibitem[{{Karlick{\'y}} {et~al.}(2012){Karlick{\'y}}, {Dzif{\v
  c}{\'a}kov{\'a}}, \& {Dud{\'{\i}}k}}]{karlicky2012}
{Karlick{\'y}}, M., {Dzif{\v c}{\'a}kov{\'a}}, E., \& {Dud{\'{\i}}k}, J. 2012,
  \aap, 537, A36

\bibitem[{{Karzas} \& {Latter}(1961)}]{KL1961}
{Karzas}, W.~J., \& {Latter}, R. 1961, \apjs, 6, 167

\bibitem[{{Kramers}(1923)}]{kramers1923}
{Kramers}, H.~A. 1923, Phi. Mag., 46, 836

\bibitem[{{Mazzotta} {et~al.}(1998){Mazzotta}, {Mazzitelli}, {Colafrancesco},
  \& {Vittorio}}]{mazzotta1998A&AS..133..403M}
{Mazzotta}, P., {Mazzitelli}, G., {Colafrancesco}, S., \& {Vittorio}, N. 1998,
  \aaps, 133, 403

\bibitem[{{Mitnik} \& {Badnell}(2004)}]{mitnik2004}
{Mitnik}, D.~M., \& {Badnell}, N.~R. 2004, \aap, 425, 1153

\bibitem[{{Mori} \& {Sugihara}(2001)}]{mori2001}
{Mori}, M., \& {Sugihara}, M. 2001, Journal of Computational and Applied
  Mathematics, 127, 287

\bibitem[{{Nahar} \& {Pradhan}(1994)}]{nahar1994PhRvA..49.1816N}
{Nahar}, S.~N., \& {Pradhan}, A.~K. 1994, \pra, 49, 1816

\bibitem[{{Nikoli{\'c}} {et~al.}(2010){Nikoli{\'c}}, {Gorczyca}, {Korista}, \&
  {Badnell}}]{nikolic2010}
{Nikoli{\'c}}, D., {Gorczyca}, T.~W., {Korista}, K.~T., \& {Badnell}, N.~R.
  2010, \aap, 516, A97

\bibitem[{{Nozawa} {et~al.}(1998){Nozawa}, {Itoh}, \& {Kohyama}}]{nozawa1998}
{Nozawa}, S., {Itoh}, N., \& {Kohyama}, Y. 1998, \apj, 507, 530

\bibitem[{{Olver} {et~al.}(2010){Olver}, {Lozier}, {Boisvert}, \&
  {Clark}}]{olver2010}
{Olver}, F.~W.~J., {Lozier}, D.~W., {Boisvert}, R.~F., \& {Clark}, C.~W. 2010,
  {NIST Handbook of Mathematical Functions} (Cambridge University Press)

\bibitem[{{Porquet} {et~al.}(2001){Porquet}, {Arnaud}, \&
  {Decourchelle}}]{porquet2001}
{Porquet}, D., {Arnaud}, M., \& {Decourchelle}, A. 2001, \aap, 373, 1110

\bibitem[{{Russell} {et~al.}(2014){Russell}, {Fabian}, {McNamara}, {Edge},
  {Sanders}, {Nulsen}, {Baum}, {Donahue}, \& {O'Dea}}]{russell2014}
{Russell}, H.~R., {Fabian}, A.~C., {McNamara}, B.~R., {et~al.} 2014, \mnras,
  444, 629

\bibitem[{{Seely} {et~al.}(1987){Seely}, {Feldman}, \& {Doschek}}]{seely1987}
{Seely}, J.~F., {Feldman}, U., \& {Doschek}, G.~A. 1987, \apj, 319, 541

\bibitem[{{Sutherland}(1998)}]{sutherland1998}
{Sutherland}, R.~S. 1998, \mnras, 300, 321

\bibitem[{{Takahasi} \& {Mori}(1974)}]{takahasi1974}
{Takahasi}, H., \& {Mori}, M. 1974, Pub. Res. Inst. Math. Sci., 9, 721

\bibitem[{{van Hoof} {et~al.}(2014){van Hoof}, {Williams}, {Volk}, {Chatzikos},
  {Ferland}, {Lykins}, {Porter}, \& {Wang}}]{vanhoof2014}
{van Hoof}, P.~A.~M., {Williams}, R.~J.~R., {Volk}, K., {et~al.} 2014, \mnras,
  444, 420

\bibitem[{{Vasyliunas}(1968)}]{vasyliunas1968}
{Vasyliunas}, V.~M. 1968, in Astrophysics and Space Science Library, Vol.~10,
  Physics of the Magnetosphere, ed. R.~D.~L. {Carovillano} \& J.~F. {McClay},
  622

\bibitem[{{Woods} {et~al.}(1981){Woods}, {Shull}, \& {Sarazin}}]{woods1981}
{Woods}, D.~T., {Shull}, J.~M., \& {Sarazin}, C.~L. 1981, \apj, 249, 399

\bibitem[{{Zatsarinny} {et~al.}(2006){Zatsarinny}, {Gorczyca}, {Fu}, {Korista},
  {Badnell}, \& {Savin}}]{zat2006}
{Zatsarinny}, O., {Gorczyca}, T.~W., {Fu}, J., {et~al.} 2006, \aap, 447, 379

\bibitem[{{Zatsarinny} {et~al.}(2003){Zatsarinny}, {Gorczyca}, {Korista},
  {Badnell}, \& {Savin}}]{zat2003}
{Zatsarinny}, O., {Gorczyca}, T.~W., {Korista}, K.~T., {Badnell}, N.~R., \&
  {Savin}, D.~W. 2003, \aap, 412, 587

\bibitem[{{Zatsarinny} {et~al.}(2004{\natexlab{a}}){Zatsarinny}, {Gorczyca},
  {Korista}, {Badnell}, \& {Savin}}]{zat2004a}
---. 2004{\natexlab{a}}, \aap, 417, 1173

\bibitem[{{Zatsarinny} {et~al.}(2004{\natexlab{b}}){Zatsarinny}, {Gorczyca},
  {Korista}, {Badnell}, \& {Savin}}]{zat2004b}
---. 2004{\natexlab{b}}, \aap, 426, 699

\bibitem[{{Zatsarinny} {et~al.}(2005{\natexlab{a}}){Zatsarinny}, {Gorczyca},
  {Korista}, {Fu}, {Badnell}, {Mitthumsiri}, \& {Savin}}]{zat2005b}
{Zatsarinny}, O., {Gorczyca}, T.~W., {Korista}, K.~T., {et~al.}
  2005{\natexlab{a}}, \aap, 440, 1203

\bibitem[{{Zatsarinny} {et~al.}(2005{\natexlab{b}}){Zatsarinny}, {Gorczyca},
  {Korista}, {Fu}, {Badnell}, {Mitthumsiri}, \& {Savin}}]{zat2005a}
---. 2005{\natexlab{b}}, \aap, 438, 743

\end{thebibliography}

\appendix

\section{The $n$-distribution}

\begin{figure}[thbp]
	\centering
	\includegraphics[width=0.45\hsize,angle=0]{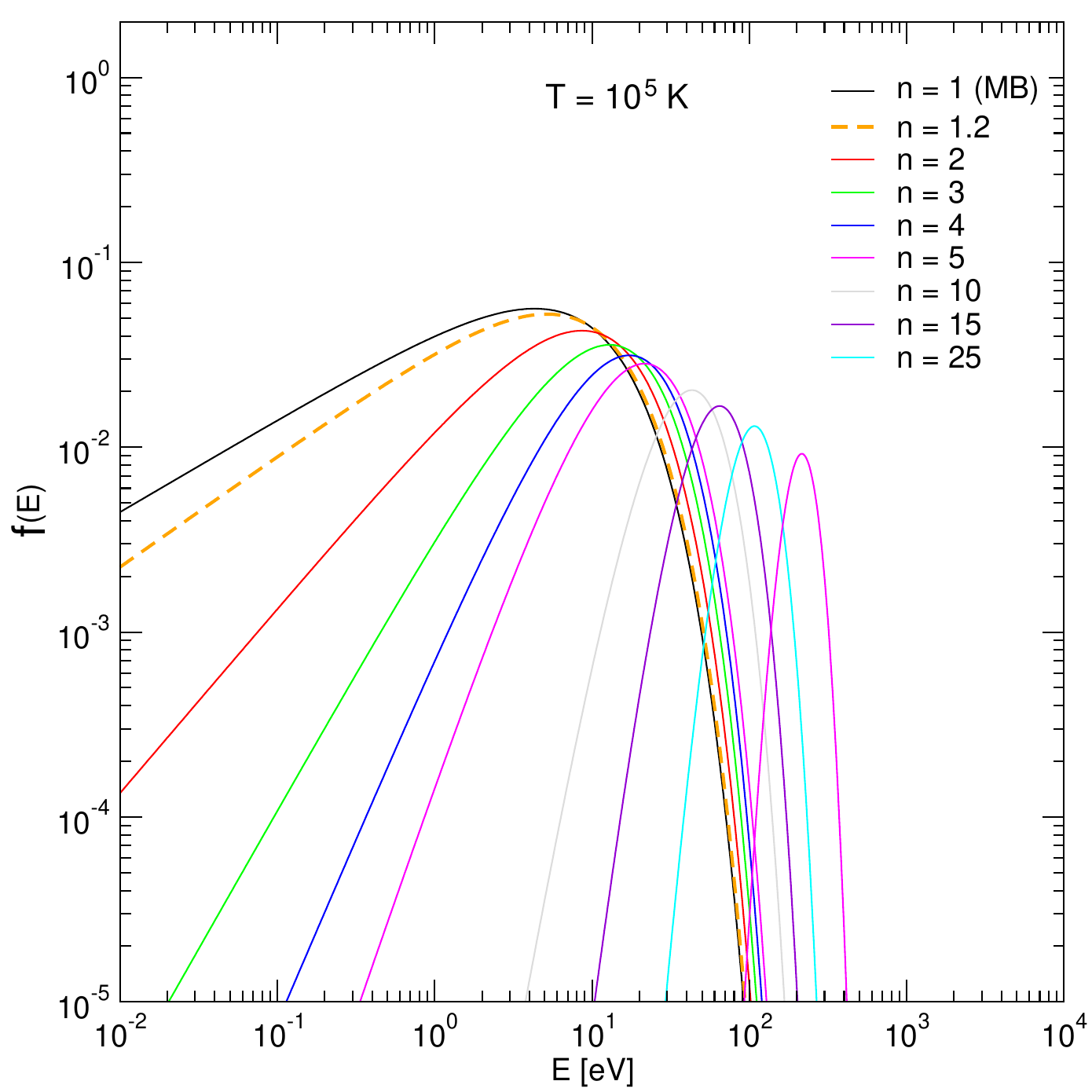}
	\includegraphics[width=0.45\hsize,angle=0]{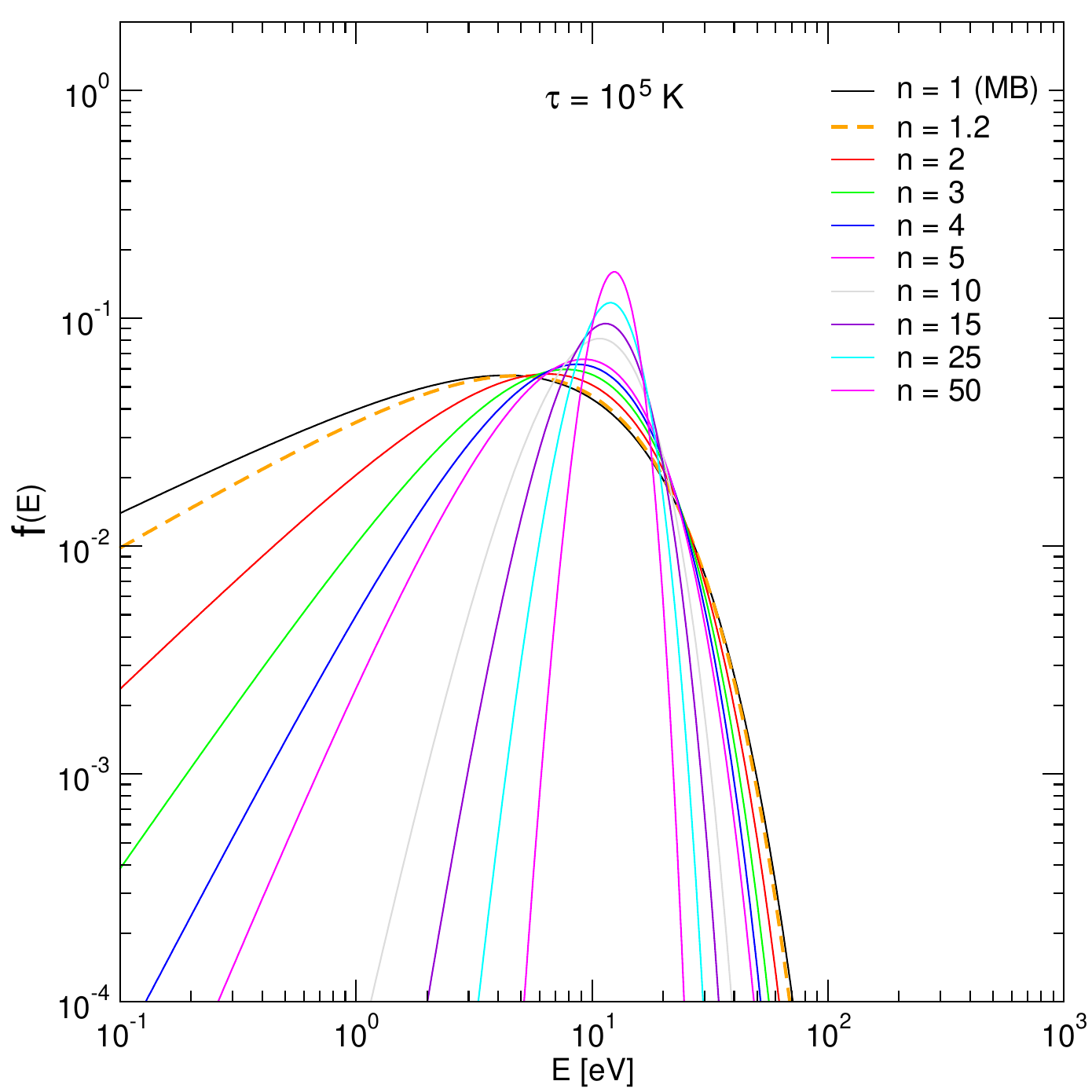}
	\caption{Maxwell-Boltzmann ($n=1$) and $n$-distributions (with $n=1.2$ through 50) at the same temperature ($T=10^{5}$ 
		K; left panel) and same pseudo-temperature ($\tau=10^{5}$ K; right panel). The $n>1$ distributions have a steeper 
		high-energy tail than the MB distribution.}
	\label{ndist}
\end{figure}

The $n-$distribution has a steeper decrease at the high-energy tail than the Maxwell-Boltzmann (MB) distribution, which in 
turn is an $n-$distribution with $n=1$. In energy space the distribution has the analytical form
\begin{equation}
f_{n}(E)dE=\frac{2}{\sqrt{\pi}(k_{B}T)^{3/2}}B_{n} E^{1/2}\left(\frac{E}{k_{B}T}\right)^{(n-1)/2}e^{-E/k_{B}T}dE,
\end{equation}
where $\displaystyle B_{n}=\frac{\sqrt{\pi}}{2\Gamma(n/2+1)}$, $E$ is the electron energy, $k_{B}$ is the 
Boltzmann 
constant, $\Gamma$ is the Gamma function, $n\in[1,+\infty[$ and $T$ are the parameters of the distribution. 
When 
$n=1$ the MB distribution is recovered. The mean energy of the distribution is given by
\begin{equation}
\langle E \rangle =\frac{3}{2}k_{B}\tau
\end{equation}
where $\displaystyle \tau=\frac{n+2}{3}T$ is the pseudo-temperature, meaning that it is the temperature of the MB distribution 
with 
the same mean energy as the mean energy of the $n-$distribution; thus, $\tau$ has the same physical meaning as $T$ in the 
MB distribution. Figure~\ref{ndist} compares the $n-$distribution for different $n$ values with the MB for the same parameter 
$T$ (top panel) and $\tau$ (bottom panel).
The $n-$distributions with the same $\tau$ have their peaks higher and narrower than that of the MB distribution, 
that is, they have less electrons with both high and low energies, but have an increased number of electrons 
with intermediate energies.

\cleardoublepage 

\section{Tables of the temperature averaged free-free Gaunt factors}

\begin{longrotatetable}	
\begin{deluxetable*}{rcccccccccccc}
		
		\tabletypesize{\scriptsize}
		\tablecaption{Temperature-averaged free-free Gaunt factor vs. $\gamma^{2}$ for different $u$ and Maxwellian 
			distribution, that is, $n$-distribution with $n=1$.}
		\tablewidth{0pt}
		\tablehead{
			\colhead{log $\gamma^{2}$} &\multicolumn{12}{c}{log $u$}\\
			\colhead{} & \colhead{-5.00} & \colhead{-4.00}& \colhead{-3.00}& \colhead{-2.00}& \colhead{-1.00}& 
			\colhead{0.00}& \colhead{1.00}& \colhead{2.00}& \colhead{3.00}& \colhead{4.00} & \colhead{5.00} & \colhead{6.00} 
		}
		\startdata
		-5.000 &  6.7989E+00 &  5.5296E+00 &  4.2615E+00 &  3.0038E+00 &  1.8090E+00 &  8.4458E-01 &  3.0462E-01 &  
		9.8499E-02 &  3.1238E-02 &  9.8827E-03 &  3.1255E-03 &  9.8838E-04 \\
		-4.900 &  6.7989E+00 &  5.5296E+00 &  4.2615E+00 &  3.0040E+00 &  1.8094E+00 &  8.4508E-01 &  3.0494E-01 &  
		9.8619E-02 &  3.1278E-02 &  9.8956E-03 &  3.1296E-03 &  9.8968E-04 \\
		-4.800 &  6.7989E+00 &  5.5296E+00 &  4.2615E+00 &  3.0042E+00 &  1.8098E+00 &  8.4563E-01 &  3.0528E-01 &  
		9.8760E-02 &  3.1326E-02 &  9.9111E-03 &  3.1345E-03 &  9.9125E-04 \\
		-4.700 &  6.7989E+00 &  5.5296E+00 &  4.2616E+00 &  3.0044E+00 &  1.8103E+00 &  8.4626E-01 &  3.0568E-01 &  
		9.8912E-02 &  3.1377E-02 &  9.9276E-03 &  3.1397E-03 &  9.9291E-04 \\
		-4.600 &  6.7989E+00 &  5.5295E+00 &  4.2616E+00 &  3.0046E+00 &  1.8108E+00 &  8.4695E-01 &  3.0611E-01 &  
		9.9090E-02 &  3.1438E-02 &  9.9471E-03 &  3.1459E-03 &  9.9487E-04 \\
		-4.500 &  6.7988E+00 &  5.5295E+00 &  4.2616E+00 &  3.0048E+00 &  1.8115E+00 &  8.4775E-01 &  3.0661E-01 &  
		9.9283E-02 &  3.1502E-02 &  9.9680E-03 &  3.1526E-03 &  9.9698E-04 \\
		-4.400 &  6.7988E+00 &  5.5295E+00 &  4.2617E+00 &  3.0051E+00 &  1.8122E+00 &  8.4863E-01 &  3.0716E-01 &  
		9.9507E-02 &  3.1578E-02 &  9.9925E-03 &  3.1604E-03 &  9.9945E-04 \\
		-4.300 &  6.7988E+00 &  5.5295E+00 &  4.2617E+00 &  3.0054E+00 &  1.8130E+00 &  8.4963E-01 &  3.0779E-01 &  
		9.9751E-02 &  3.1661E-02 &  1.0019E-02 &  3.1688E-03 &  1.0021E-03 \\
\enddata
	\end{deluxetable*}
\end{longrotatetable}
	
	\clearpage

\begin{longrotatetable}	
	\begin{deluxetable*}{rcccccccccccc}
		\tabletypesize{\scriptsize}
		\tablecaption{Temperature-averaged free-free Gaunt factor vs. $\gamma^{2}$ for different $u$ and 
			$n$-distribution: $n=1.2$.}
		\tablewidth{0pt}
		\tablehead{
			\colhead{log $\gamma^{2}$} &\multicolumn{12}{c}{log $u$}\\
			\colhead{} & \colhead{-5.00} & \colhead{-4.00}& \colhead{-3.00}& \colhead{-2.00}& \colhead{-1.00}& 
			\colhead{0.00}& \colhead{1.00}& \colhead{2.00}& \colhead{3.00}& \colhead{4.00} & \colhead{5.00} & \colhead{6.00} 
		}
		\startdata
		-5.000 &  6.8832E+00 &  5.6130E+00 &  4.3431E+00 &  3.0787E+00 &  1.8603E+00 &  8.5977E-01 &  3.0587E-01 &  
		9.8547E-02 &  3.1240E-02 &  9.8827E-03 &  3.1255E-03 &  9.8838E-04 \\
		-4.900 &  6.8832E+00 &  5.6130E+00 &  4.3431E+00 &  3.0789E+00 &  1.8607E+00 &  8.6027E-01 &  3.0619E-01 &  
		9.8666E-02 &  3.1280E-02 &  9.8956E-03 &  3.1296E-03 &  9.8968E-04 \\
		-4.800 &  6.8832E+00 &  5.6130E+00 &  4.3432E+00 &  3.0790E+00 &  1.8611E+00 &  8.6080E-01 &  3.0653E-01 &  
		9.8808E-02 &  3.1328E-02 &  9.9112E-03 &  3.1345E-03 &  9.9125E-04 \\
		-4.700 &  6.8832E+00 &  5.6130E+00 &  4.3432E+00 &  3.0792E+00 &  1.8615E+00 &  8.6142E-01 &  3.0693E-01 &  
		9.8960E-02 &  3.1379E-02 &  9.9276E-03 &  3.1397E-03 &  9.9291E-04 \\
		-4.600 &  6.8832E+00 &  5.6130E+00 &  4.3432E+00 &  3.0793E+00 &  1.8620E+00 &  8.6210E-01 &  3.0736E-01 &  
		9.9138E-02 &  3.1439E-02 &  9.9471E-03 &  3.1459E-03 &  9.9487E-04 \\
		-4.500 &  6.8832E+00 &  5.6130E+00 &  4.3432E+00 &  3.0795E+00 &  1.8626E+00 &  8.6288E-01 &  3.0786E-01 &  
		9.9330E-02 &  3.1504E-02 &  9.9680E-03 &  3.1526E-03 &  9.9698E-04 \\
		-4.400 &  6.8832E+00 &  5.6130E+00 &  4.3433E+00 &  3.0798E+00 &  1.8632E+00 &  8.6373E-01 &  3.0841E-01 &  
		9.9555E-02 &  3.1580E-02 &  9.9925E-03 &  3.1604E-03 &  9.9945E-04 \\
		-4.300 &  6.8831E+00 &  5.6130E+00 &  4.3433E+00 &  3.0800E+00 &  1.8639E+00 &  8.6471E-01 &  3.0904E-01 &  
		9.9799E-02 &  3.1662E-02 &  1.0019E-02 &  3.1689E-03 &  1.0021E-03 \\
\enddata
	\end{deluxetable*}
\end{longrotatetable}

	\clearpage
	
	\begin{longrotatetable}	
		\begin{deluxetable*}{rcccccccccccc}
		\tabletypesize{\scriptsize}
		\tablecaption{Temperature-averaged free-free Gaunt factor vs. $\gamma^{2}$ for different $u$ and 
			$n$-distribution: $n=2$.}
		\tablewidth{0pt}
		\tablehead{
			\colhead{log $\gamma^{2}$} &\multicolumn{12}{c}{log $u$}\\
			\colhead{} & \colhead{-5.00} & \colhead{-4.00}& \colhead{-3.00}& \colhead{-2.00}& \colhead{-1.00}& 
			\colhead{0.00}& \colhead{1.00}& \colhead{2.00}& \colhead{3.00}& \colhead{4.00} & \colhead{5.00} & \colhead{6.00} 
		}
		\startdata
		-5.000 &  7.1328E+00 &  5.8630E+00 &  4.5929E+00 &  3.3210E+00 &  2.0508E+00 &  9.2375E-01 &  3.1100E-01 &  
		9.8738E-02 &  3.1246E-02 &  9.8829E-03 &  3.1255E-03 &  9.8838E-04 \\
		-4.900 &  7.1328E+00 &  5.8630E+00 &  4.5929E+00 &  3.3211E+00 &  2.0510E+00 &  9.2420E-01 &  3.1131E-01 &  
		9.8857E-02 &  3.1286E-02 &  9.8958E-03 &  3.1296E-03 &  9.8968E-04 \\
		-4.800 &  7.1328E+00 &  5.8630E+00 &  4.5929E+00 &  3.3212E+00 &  2.0513E+00 &  9.2469E-01 &  3.1166E-01 &  
		9.8999E-02 &  3.1334E-02 &  9.9114E-03 &  3.1345E-03 &  9.9125E-04 \\
		-4.700 &  7.1328E+00 &  5.8630E+00 &  4.5929E+00 &  3.3212E+00 &  2.0516E+00 &  9.2525E-01 &  3.1205E-01 &  
		9.9151E-02 &  3.1385E-02 &  9.9278E-03 &  3.1397E-03 &  9.9291E-04 \\
		-4.600 &  7.1328E+00 &  5.8630E+00 &  4.5929E+00 &  3.3213E+00 &  2.0519E+00 &  9.2587E-01 &  3.1248E-01 &  
		9.9329E-02 &  3.1445E-02 &  9.9473E-03 &  3.1460E-03 &  9.9487E-04 \\
		-4.500 &  7.1328E+00 &  5.8630E+00 &  4.5929E+00 &  3.3214E+00 &  2.0522E+00 &  9.2658E-01 &  3.1298E-01 &  
		9.9521E-02 &  3.1510E-02 &  9.9682E-03 &  3.1526E-03 &  9.9698E-04 \\
		-4.400 &  7.1328E+00 &  5.8630E+00 &  4.5929E+00 &  3.3215E+00 &  2.0527E+00 &  9.2735E-01 &  3.1352E-01 &  
		9.9746E-02 &  3.1586E-02 &  9.9927E-03 &  3.1604E-03 &  9.9945E-04 \\
		-4.300 &  7.1328E+00 &  5.8630E+00 &  4.5929E+00 &  3.3216E+00 &  2.0531E+00 &  9.2824E-01 &  3.1415E-01 &  
		9.9990E-02 &  3.1668E-02 &  1.0019E-02 &  3.1689E-03 &  1.0021E-03 \\
\enddata
\end{deluxetable*}
\end{longrotatetable}
	\clearpage
	
	\begin{longrotatetable}	
		\begin{deluxetable*}{rcccccccccccc}
		\tabletypesize{\scriptsize}
		\tablecaption{Temperature-averaged free-free Gaunt factor vs. $\gamma^{2}$ for different $u$ and 
			$n$-distribution: $n=3$.}
		\tablewidth{0pt}
		\tablehead{
			\colhead{log $\gamma^{2}$} &\multicolumn{12}{c}{log $u$}\\
			\colhead{} & \colhead{-5.00} & \colhead{-4.00}& \colhead{-3.00}& \colhead{-2.00}& \colhead{-1.00}& 
			\colhead{0.00}& \colhead{1.00}& \colhead{2.00}& \colhead{3.00}& \colhead{4.00} & \colhead{5.00} & \colhead{6.00} 
		}
		\startdata
		-5.000 &  7.3448E+00 &  6.0753E+00 &  4.8056E+00 &  3.5339E+00 &  2.2486E+00 &  1.0099E+00 &  3.1773E-01 &  
		9.8979E-02 &  3.1253E-02 &  9.8832E-03 &  3.1255E-03 &  9.8838E-04 \\
		-4.900 &  7.3448E+00 &  6.0753E+00 &  4.8056E+00 &  3.5339E+00 &  2.2488E+00 &  1.0103E+00 &  3.1804E-01 &  
		9.9098E-02 &  3.1294E-02 &  9.8961E-03 &  3.1296E-03 &  9.8968E-04 \\
		-4.800 &  7.3448E+00 &  6.0753E+00 &  4.8056E+00 &  3.5340E+00 &  2.2489E+00 &  1.0107E+00 &  3.1838E-01 &  
		9.9240E-02 &  3.1342E-02 &  9.9116E-03 &  3.1345E-03 &  9.9125E-04 \\
		-4.700 &  7.3448E+00 &  6.0753E+00 &  4.8056E+00 &  3.5340E+00 &  2.2491E+00 &  1.0112E+00 &  3.1877E-01 &  
		9.9391E-02 &  3.1393E-02 &  9.9280E-03 &  3.1398E-03 &  9.9291E-04 \\
		-4.600 &  7.3448E+00 &  6.0753E+00 &  4.8056E+00 &  3.5340E+00 &  2.2493E+00 &  1.0118E+00 &  3.1919E-01 &  
		9.9569E-02 &  3.1453E-02 &  9.9476E-03 &  3.1460E-03 &  9.9487E-04 \\
		-4.500 &  7.3447E+00 &  6.0752E+00 &  4.8056E+00 &  3.5341E+00 &  2.2495E+00 &  1.0124E+00 &  3.1968E-01 &  
		9.9762E-02 &  3.1518E-02 &  9.9684E-03 &  3.1526E-03 &  9.9698E-04 \\
		-4.400 &  7.3447E+00 &  6.0752E+00 &  4.8056E+00 &  3.5341E+00 &  2.2497E+00 &  1.0131E+00 &  3.2023E-01 &  
		9.9986E-02 &  3.1594E-02 &  9.9930E-03 &  3.1604E-03 &  9.9945E-04 \\
		-4.300 &  7.3447E+00 &  6.0752E+00 &  4.8056E+00 &  3.5342E+00 &  2.2500E+00 &  1.0138E+00 &  3.2085E-01 &  
		1.0023E-01 &  3.1676E-02 &  1.0020E-02 &  3.1689E-03 &  1.0021E-03 \\
\enddata
	\end{deluxetable*}
\end{longrotatetable}

	\clearpage
	
	\begin{longrotatetable}	
		\begin{deluxetable*}{rcccccccccccc}
		\tabletypesize{\scriptsize}
		\tablecaption{Temperature-averaged free-free Gaunt factor vs. $\gamma^{2}$ for different $u$ and 
			$n$-distribution: $n=5$.}
		\tablewidth{0pt}
		\tablehead{
			\colhead{log $\gamma^{2}$} &\multicolumn{12}{c}{log $u$}\\
			\colhead{} & \colhead{-5.00} & \colhead{-4.00}& \colhead{-3.00}& \colhead{-2.00}& \colhead{-1.00}& 
			\colhead{0.00}& \colhead{1.00}& \colhead{2.00}& \colhead{3.00}& \colhead{4.00} & \colhead{5.00} & \colhead{6.00} 
		}
		\startdata
		-5.000 &  7.6205E+00 &  6.3510E+00 &  5.0814E+00 &  3.8107E+00 &  2.5291E+00 &  1.1928E+00 &  3.3232E-01 &  
		9.9465E-02 &  3.1269E-02 &  9.8836E-03 &  3.1255E-03 &  9.8838E-04 \\
		-4.900 &  7.6205E+00 &  6.3510E+00 &  5.0814E+00 &  3.8107E+00 &  2.5291E+00 &  1.1931E+00 &  3.3262E-01 &  
		9.9584E-02 &  3.1309E-02 &  9.8966E-03 &  3.1296E-03 &  9.8968E-04 \\
		-4.800 &  7.6205E+00 &  6.3510E+00 &  5.0814E+00 &  3.8107E+00 &  2.5292E+00 &  1.1934E+00 &  3.3296E-01 &  
		9.9726E-02 &  3.1357E-02 &  9.9121E-03 &  3.1345E-03 &  9.9125E-04 \\
		-4.700 &  7.6205E+00 &  6.3510E+00 &  5.0814E+00 &  3.8107E+00 &  2.5292E+00 &  1.1938E+00 &  3.3334E-01 &  
		9.9877E-02 &  3.1408E-02 &  9.9285E-03 &  3.1398E-03 &  9.9291E-04 \\
		-4.600 &  7.6205E+00 &  6.3510E+00 &  5.0814E+00 &  3.8107E+00 &  2.5293E+00 &  1.1942E+00 &  3.3376E-01 &  
		1.0006E-01 &  3.1469E-02 &  9.9480E-03 &  3.1460E-03 &  9.9487E-04 \\
		-4.500 &  7.6205E+00 &  6.3510E+00 &  5.0814E+00 &  3.8108E+00 &  2.5294E+00 &  1.1946E+00 &  3.3424E-01 &  
		1.0025E-01 &  3.1533E-02 &  9.9689E-03 &  3.1526E-03 &  9.9698E-04 \\
		-4.400 &  7.6205E+00 &  6.3510E+00 &  5.0814E+00 &  3.8108E+00 &  2.5295E+00 &  1.1951E+00 &  3.3478E-01 &  
		1.0047E-01 &  3.1609E-02 &  9.9935E-03 &  3.1604E-03 &  9.9945E-04 \\
		-4.300 &  7.6205E+00 &  6.3510E+00 &  5.0814E+00 &  3.8108E+00 &  2.5296E+00 &  1.1957E+00 &  3.3538E-01 &  
		1.0072E-01 &  3.1692E-02 &  1.0020E-02 &  3.1689E-03 &  1.0021E-03 \\
\enddata
	\end{deluxetable*}
\end{longrotatetable}

	\clearpage
	
	\begin{longrotatetable}	
		\begin{deluxetable*}{rcccccccccccc}
		\tabletypesize{\scriptsize}
		\tablecaption{Temperature-averaged free-free Gaunt factor vs. $\gamma^{2}$ for different $u$ and 
			$n$-distribution: $n=10$.}
		\tablewidth{0pt}
		\tablehead{
			\colhead{log $\gamma^{2}$} &\multicolumn{12}{c}{log $u$}\\
			\colhead{} & \colhead{-5.00} & \colhead{-4.00}& \colhead{-3.00}& \colhead{-2.00}& \colhead{-1.00}& 
			\colhead{0.00}& \colhead{1.00}& \colhead{2.00}& \colhead{3.00}& \colhead{4.00} & \colhead{5.00} & \colhead{6.00} 
		}
		\startdata
		-5.000 &  8.0000E+00 &  6.7305E+00 &  5.4609E+00 &  4.1909E+00 &  2.9160E+00 &  1.5859E+00 &  3.7691E-01 &  
		1.0071E-01 &  3.1308E-02 &  9.8849E-03 &  3.1255E-03 &  9.8838E-04 \\
		-4.900 &  8.0000E+00 &  6.7305E+00 &  5.4609E+00 &  4.1909E+00 &  2.9160E+00 &  1.5861E+00 &  3.7720E-01 &  
		1.0083E-01 &  3.1348E-02 &  9.8978E-03 &  3.1296E-03 &  9.8968E-04 \\
		-4.800 &  8.0000E+00 &  6.7305E+00 &  5.4609E+00 &  4.1909E+00 &  2.9160E+00 &  1.5862E+00 &  3.7751E-01 &  
		1.0097E-01 &  3.1396E-02 &  9.9133E-03 &  3.1346E-03 &  9.9125E-04 \\
		-4.700 &  8.0000E+00 &  6.7305E+00 &  5.4609E+00 &  4.1909E+00 &  2.9160E+00 &  1.5863E+00 &  3.7788E-01 &  
		1.0112E-01 &  3.1447E-02 &  9.9298E-03 &  3.1398E-03 &  9.9291E-04 \\
		-4.600 &  8.0000E+00 &  6.7305E+00 &  5.4609E+00 &  4.1909E+00 &  2.9161E+00 &  1.5865E+00 &  3.7827E-01 &  
		1.0130E-01 &  3.1507E-02 &  9.9493E-03 &  3.1460E-03 &  9.9487E-04 \\
		-4.500 &  8.0000E+00 &  6.7305E+00 &  5.4609E+00 &  4.1909E+00 &  2.9161E+00 &  1.5867E+00 &  3.7873E-01 &  
		1.0149E-01 &  3.1572E-02 &  9.9702E-03 &  3.1527E-03 &  9.9698E-04 \\
		-4.400 &  7.9999E+00 &  6.7305E+00 &  5.4609E+00 &  4.1909E+00 &  2.9161E+00 &  1.5869E+00 &  3.7923E-01 &  
		1.0172E-01 &  3.1648E-02 &  9.9947E-03 &  3.1605E-03 &  9.9945E-04 \\
		-4.300 &  7.9999E+00 &  6.7305E+00 &  5.4609E+00 &  4.1909E+00 &  2.9162E+00 &  1.5872E+00 &  3.7981E-01 &  
		1.0196E-01 &  3.1730E-02 &  1.0021E-02 &  3.1689E-03 &  1.0021E-03 \\
\enddata
	\end{deluxetable*}
\end{longrotatetable}
	\clearpage
	
	\begin{longrotatetable}	
		\begin{deluxetable*}{rcccccccccccc}
		\tabletypesize{\scriptsize}
		\tablecaption{Temperature-averaged free-free Gaunt factor vs. $\gamma^{2}$ for different $u$ and 
			$n$-distribution: $n=15$.}
		\tablewidth{0pt}
		\tablehead{
			\colhead{log $\gamma^{2}$} &\multicolumn{12}{c}{log $u$}\\
			\colhead{} & \colhead{-5.00} & \colhead{-4.00}& \colhead{-3.00}& \colhead{-2.00}& \colhead{-1.00}& 
			\colhead{0.00}& \colhead{1.00}& \colhead{2.00}& \colhead{3.00}& \colhead{4.00} & \colhead{5.00} & \colhead{6.00} 
		}
		\startdata
		-5.000 &  8.2230E+00 &  6.9535E+00 &  5.6840E+00 &  4.4142E+00 &  3.1412E+00 &  1.8340E+00 &  4.3677E-01 &  
		1.0201E-01 &  3.1346E-02 &  9.8861E-03 &  3.1256E-03 &  9.8838E-04 \\
		-4.900 &  8.2230E+00 &  6.9535E+00 &  5.6840E+00 &  4.4142E+00 &  3.1412E+00 &  1.8340E+00 &  4.3704E-01 &  
		1.0213E-01 &  3.1386E-02 &  9.8990E-03 &  3.1297E-03 &  9.8968E-04 \\
		-4.800 &  8.2230E+00 &  6.9535E+00 &  5.6840E+00 &  4.4142E+00 &  3.1412E+00 &  1.8341E+00 &  4.3733E-01 &  
		1.0227E-01 &  3.1435E-02 &  9.9145E-03 &  3.1346E-03 &  9.9125E-04 \\
		-4.700 &  8.2230E+00 &  6.9535E+00 &  5.6840E+00 &  4.4142E+00 &  3.1412E+00 &  1.8342E+00 &  4.3767E-01 &  
		1.0242E-01 &  3.1486E-02 &  9.9310E-03 &  3.1398E-03 &  9.9291E-04 \\
		-4.600 &  8.2230E+00 &  6.9535E+00 &  5.6840E+00 &  4.4142E+00 &  3.1412E+00 &  1.8343E+00 &  4.3803E-01 &  
		1.0260E-01 &  3.1546E-02 &  9.9505E-03 &  3.1461E-03 &  9.9488E-04 \\
		-4.500 &  8.2230E+00 &  6.9535E+00 &  5.6840E+00 &  4.4142E+00 &  3.1412E+00 &  1.8344E+00 &  4.3846E-01 &  
		1.0279E-01 &  3.1611E-02 &  9.9714E-03 &  3.1527E-03 &  9.9698E-04 \\
		-4.400 &  8.2230E+00 &  6.9535E+00 &  5.6840E+00 &  4.4142E+00 &  3.1413E+00 &  1.8345E+00 &  4.3892E-01 &  
		1.0301E-01 &  3.1687E-02 &  9.9959E-03 &  3.1605E-03 &  9.9946E-04 \\
		-4.300 &  8.2230E+00 &  6.9535E+00 &  5.6840E+00 &  4.4142E+00 &  3.1413E+00 &  1.8346E+00 &  4.3946E-01 &  
		1.0326E-01 &  3.1769E-02 &  1.0022E-02 &  3.1690E-03 &  1.0021E-03 \\
\enddata
	\end{deluxetable*}
\end{longrotatetable}

	\clearpage
	
	\begin{longrotatetable}	
		\begin{deluxetable*}{rcccccccccccc}
		\tabletypesize{\scriptsize}
		\tablecaption{Temperature-averaged free-free Gaunt factor vs. $\gamma^{2}$ for different $u$ and 
			$n$-distribution: $n=25$.}
		\tablewidth{0pt}
		\tablehead{
			\colhead{log $\gamma^{2}$} &\multicolumn{12}{c}{log $u$}\\
			\colhead{} & \colhead{-5.00} & \colhead{-4.00}& \colhead{-3.00}& \colhead{-2.00}& \colhead{-1.00}& 
			\colhead{0.00}& \colhead{1.00}& \colhead{2.00}& \colhead{3.00}& \colhead{4.00} & \colhead{5.00} & \colhead{6.00} 
		}
		\startdata
		-5.000 &  8.5044E+00 &  7.2349E+00 &  5.9654E+00 &  4.6957E+00 &  3.4242E+00 &  2.1334E+00 &  6.1861E-01 &  
		1.0475E-01 &  3.1425E-02 &  9.8885E-03 &  3.1256E-03 &  9.8839E-04 \\
		-4.900 &  8.5044E+00 &  7.2349E+00 &  5.9654E+00 &  4.6957E+00 &  3.4242E+00 &  2.1334E+00 &  6.1884E-01 &  
		1.0487E-01 &  3.1465E-02 &  9.9014E-03 &  3.1298E-03 &  9.8969E-04 \\
		-4.800 &  8.5044E+00 &  7.2349E+00 &  5.9654E+00 &  4.6957E+00 &  3.4242E+00 &  2.1335E+00 &  6.1904E-01 &  
		1.0501E-01 &  3.1513E-02 &  9.9170E-03 &  3.1347E-03 &  9.9125E-04 \\
		-4.700 &  8.5044E+00 &  7.2349E+00 &  5.9654E+00 &  4.6957E+00 &  3.4242E+00 &  2.1335E+00 &  6.1932E-01 &  
		1.0516E-01 &  3.1564E-02 &  9.9334E-03 &  3.1399E-03 &  9.9291E-04 \\
		-4.600 &  8.5044E+00 &  7.2349E+00 &  5.9654E+00 &  4.6957E+00 &  3.4242E+00 &  2.1336E+00 &  6.1959E-01 &  
		1.0534E-01 &  3.1624E-02 &  9.9529E-03 &  3.1461E-03 &  9.9488E-04 \\
		-4.500 &  8.5044E+00 &  7.2349E+00 &  5.9654E+00 &  4.6957E+00 &  3.4242E+00 &  2.1336E+00 &  6.1994E-01 &  
		1.0553E-01 &  3.1689E-02 &  9.9738E-03 &  3.1528E-03 &  9.9698E-04 \\
		-4.400 &  8.5044E+00 &  7.2349E+00 &  5.9654E+00 &  4.6957E+00 &  3.4242E+00 &  2.1337E+00 &  6.2028E-01 &  
		1.0575E-01 &  3.1765E-02 &  9.9984E-03 &  3.1606E-03 &  9.9946E-04 \\
		-4.300 &  8.5044E+00 &  7.2349E+00 &  5.9654E+00 &  4.6957E+00 &  3.4242E+00 &  2.1337E+00 &  6.2072E-01 &  
		1.0600E-01 &  3.1847E-02 &  1.0025E-02 &  3.1690E-03 &  1.0021E-03 \\
\enddata
	\end{deluxetable*}
\end{longrotatetable}

	\clearpage
	
	\begin{longrotatetable}	
		\begin{deluxetable*}{rcccccccccccc}
		\tabletypesize{\scriptsize}
		\tablecaption{Temperature-averaged free-free Gaunt factor vs. $\gamma^{2}$ for different $u$ and 
			$n$-distribution: $n=50$.}
		\tablewidth{0pt}
		\tablehead{
			\colhead{log $\gamma^{2}$} &\multicolumn{12}{c}{log $u$}\\
			\colhead{} & \colhead{-5.00} & \colhead{-4.00}& \colhead{-3.00}& \colhead{-2.00}& \colhead{-1.00}& 
			\colhead{0.00}& \colhead{1.00}& \colhead{2.00}& \colhead{3.00}& \colhead{4.00} & \colhead{5.00} & \colhead{6.00} 
		}
		\startdata
		-5.000 &  8.8864E+00 &  7.6169E+00 &  6.3474E+00 &  5.0779E+00 &  3.8074E+00 &  2.5277E+00 &  1.1332E+00 &  
		1.1270E-01 &  3.1622E-02 &  9.8947E-03 &  3.1258E-03 &  9.8839E-04 \\
		-4.900 &  8.8864E+00 &  7.6169E+00 &  6.3474E+00 &  5.0779E+00 &  3.8074E+00 &  2.5277E+00 &  1.1333E+00 &  
		1.1282E-01 &  3.1662E-02 &  9.9076E-03 &  3.1299E-03 &  9.8969E-04 \\
		-4.800 &  8.8864E+00 &  7.6169E+00 &  6.3474E+00 &  5.0779E+00 &  3.8074E+00 &  2.5277E+00 &  1.1333E+00 &  
		1.1296E-01 &  3.1711E-02 &  9.9231E-03 &  3.1349E-03 &  9.9126E-04 \\
		-4.700 &  8.8864E+00 &  7.6169E+00 &  6.3474E+00 &  5.0779E+00 &  3.8074E+00 &  2.5277E+00 &  1.1335E+00 &  
		1.1310E-01 &  3.1761E-02 &  9.9395E-03 &  3.1401E-03 &  9.9292E-04 \\
		-4.600 &  8.8864E+00 &  7.6169E+00 &  6.3474E+00 &  5.0779E+00 &  3.8074E+00 &  2.5277E+00 &  1.1336E+00 &  
		1.1328E-01 &  3.1822E-02 &  9.9590E-03 &  3.1463E-03 &  9.9488E-04 \\
		-4.500 &  8.8864E+00 &  7.6169E+00 &  6.3474E+00 &  5.0779E+00 &  3.8074E+00 &  2.5277E+00 &  1.1337E+00 &  
		1.1347E-01 &  3.1887E-02 &  9.9799E-03 &  3.1530E-03 &  9.9699E-04 \\
		-4.400 &  8.8864E+00 &  7.6169E+00 &  6.3474E+00 &  5.0779E+00 &  3.8074E+00 &  2.5278E+00 &  1.1338E+00 &  
		1.1369E-01 &  3.1963E-02 &  1.0004E-02 &  3.1608E-03 &  9.9946E-04 \\
		-4.300 &  8.8864E+00 &  7.6169E+00 &  6.3474E+00 &  5.0779E+00 &  3.8074E+00 &  2.5278E+00 &  1.1340E+00 &  
		1.1393E-01 &  3.2045E-02 &  1.0031E-02 &  3.1692E-03 &  1.0021E-03 \\
\enddata
	\end{deluxetable*}
\end{longrotatetable}
		
	\clearpage
	
	\begin{longrotatetable}	
		\begin{deluxetable*}{rcccccccccccc}
		\tabletypesize{\scriptsize}
		\tablecaption{Temperature-averaged free-free Gaunt factor vs. $\gamma^{2}$ for different $u$ and 
			$n$-distribution: $n=100$.}
		\tablewidth{0pt}
		\tablehead{
			\colhead{log $\gamma^{2}$} &\multicolumn{12}{c}{log $u$}\\
			\colhead{} & \colhead{-5.00} & \colhead{-4.00}& \colhead{-3.00}& \colhead{-2.00}& \colhead{-1.00}& 
			\colhead{0.00}& \colhead{1.00}& \colhead{2.00}& \colhead{3.00}& \colhead{4.00} & \colhead{5.00} & \colhead{6.00} 
		}
		\startdata
		-5.000 &  9.2685E+00 &  7.9990E+00 &  6.7296E+00 &  5.4600E+00 &  4.1901E+00 &  2.9155E+00 &  1.5913E+00 &  
		1.3577E-01 &  3.2030E-02 &  9.9069E-03 &  3.1262E-03 &  9.8840E-04 \\
		-4.900 &  9.2685E+00 &  7.9990E+00 &  6.7296E+00 &  5.4600E+00 &  4.1901E+00 &  2.9155E+00 &  1.5913E+00 &  
		1.3589E-01 &  3.2070E-02 &  9.9198E-03 &  3.1303E-03 &  9.8971E-04 \\
		-4.800 &  9.2685E+00 &  7.9990E+00 &  6.7296E+00 &  5.4600E+00 &  4.1901E+00 &  2.9156E+00 &  1.5914E+00 &  
		1.3602E-01 &  3.2118E-02 &  9.9354E-03 &  3.1353E-03 &  9.9127E-04 \\
		-4.700 &  9.2685E+00 &  7.9990E+00 &  6.7296E+00 &  5.4600E+00 &  4.1901E+00 &  2.9156E+00 &  1.5914E+00 &  
		1.3617E-01 &  3.2169E-02 &  9.9518E-03 &  3.1405E-03 &  9.9293E-04 \\
		-4.600 &  9.2685E+00 &  7.9990E+00 &  6.7296E+00 &  5.4600E+00 &  4.1901E+00 &  2.9156E+00 &  1.5915E+00 &  
		1.3634E-01 &  3.2229E-02 &  9.9713E-03 &  3.1467E-03 &  9.9490E-04 \\
		-4.500 &  9.2685E+00 &  7.9990E+00 &  6.7296E+00 &  5.4600E+00 &  4.1901E+00 &  2.9156E+00 &  1.5915E+00 &  
		1.3652E-01 &  3.2294E-02 &  9.9922E-03 &  3.1534E-03 &  9.9700E-04 \\
		-4.400 &  9.2685E+00 &  7.9990E+00 &  6.7296E+00 &  5.4600E+00 &  4.1901E+00 &  2.9156E+00 &  1.5916E+00 &  
		1.3674E-01 &  3.2370E-02 &  1.0017E-02 &  3.1612E-03 &  9.9948E-04 \\
		-4.300 &  9.2685E+00 &  7.9990E+00 &  6.7296E+00 &  5.4600E+00 &  4.1901E+00 &  2.9156E+00 &  1.5916E+00 &  
		1.3697E-01 &  3.2452E-02 &  1.0043E-02 &  3.1696E-03 &  1.0022E-03 \\
\enddata
\end{deluxetable*}
\end{longrotatetable}

\cleardoublepage
	
\section{Table of the total free-free Gaunt factor for different $n$}

	\begin{longrotatetable}	
	\begin{deluxetable*}{rcccccccccc}
		\tabletypesize{\scriptsize}
		\tablecaption{Total free-free Gaunt factor for the $n$-distribution of electron energies with $n$ varying 
			between 1 and 100.\label{tot_gff}}
		\tablewidth{0pt}
		\tablehead{
			\colhead{log $\gamma^{2}$} &\multicolumn{10}{c}{$n$}\\
			\colhead{} & \colhead{1}& \colhead{1.2}& \colhead{2}& \colhead{3}& \colhead{5}& 
			\colhead{10}& \colhead{15}& \colhead{25}& \colhead{50}& \colhead{100}
		}
		\startdata 
		-5.000 &  1.1041E+00 &  1.1460E+00 &  1.3009E+00 &  1.4723E+00 &  1.7667E+00 &  2.3471E+00 &  2.8104E+00 &  
		3.5606E+00 &  4.9624E+00 &  6.9653E+00 \\
		-4.900 &  1.1045E+00 &  1.1465E+00 &  1.3013E+00 &  1.4728E+00 &  1.7671E+00 &  2.3476E+00 &  2.8109E+00 &  
		3.5610E+00 
		&  4.9628E+00 &  6.9658E+00 \\
		-4.800 &  1.1051E+00 &  1.1470E+00 &  1.3018E+00 &  1.4733E+00 &  1.7676E+00 &  2.3481E+00 &  2.8114E+00 &  
		3.5615E+00 
		&  4.9633E+00 &  6.9663E+00 \\
		-4.700 &  1.1056E+00 &  1.1476E+00 &  1.3024E+00 &  1.4739E+00 &  1.7682E+00 &  2.3487E+00 &  2.8120E+00 &  
		3.5621E+00 &  4.9639E+00 &  6.9669E+00 \\
		-4.600 &  1.1063E+00 &  1.1482E+00 &  1.3031E+00 &  1.4745E+00 &  1.7689E+00 &  2.3493E+00 &  2.8126E+00 &  
		3.5628E+00 &  4.9646E+00 &  6.9675E+00 \\
		-4.500 &  1.1070E+00 &  1.1489E+00 &  1.3038E+00 &  1.4752E+00 &  1.7696E+00 &  2.3501E+00 &  2.8134E+00 &  
		3.5635E+00 &  4.9653E+00 &  6.9683E+00 \\
		-4.400 &  1.1078E+00 &  1.1498E+00 &  1.3046E+00 &  1.4761E+00 &  1.7704E+00 &  2.3509E+00 &  2.8142E+00 &  
		3.5643E+00 &  4.9661E+00 &  6.9691E+00 \\
		-4.300 &  1.1087E+00 &  1.1507E+00 &  1.3055E+00 &  1.4770E+00 &  1.7713E+00 &  2.3518E+00 &  2.8151E+00 &  
		3.5652E+00 
		&  4.9670E+00 &  6.9700E+00 \\
\enddata
	\end{deluxetable*}
\end{longrotatetable}

\end{document}